\documentstyle[draft,eqsecnum,aps]{revtex}    
\begin{document}

\draft

\title{Quantum interference and the spin orbit interaction \\ in  \\
mesoscopic normal-superconducting junctions} 

\author{Keith Slevin and Jean-Louis Pichard}

\address{Service de Physique de l'\'Etat Condens\'e, CEA-Saclay, 91191
Gif-sur-Yvette,\\ France}

\author{Pier A. Mello}

\address{Instituto de Fisica, UNAM, Apartado Postal 20-364, 01000
Mexico D.F.}

\maketitle
\bigskip

\noindent
Short title: The SO interaction in
mesoscopic NS junctions.
\bigskip

\begin{abstract}
We calculate the quantum correction to the classical conductance
of a disordered mesoscopic normal-superconducting (NS) junction in 
which the electron 
spatial and spin degrees of freedom are coupled by an appreciable
spin orbit interaction.
We use random matrix theory to describe the scattering in the normal
part of the junction and consider both quasi-ballistic and 
diffusive junctions. 
The dependence of the junction conductance on the 
Schottky barrier transparency 
at the NS interface is also considered.
We find that the quantum correction is sensitive to the breaking
of spin rotation symmetry even when the junction is in a 
magnetic field and time reversal symmetry is broken.
We demonstrate that this sensitivity is due to quantum interference
between scattering processes which involve electrons and holes
traversing closed loops in the same direction.
We explain why such processes are sensitive to the spin orbit
interaction but not to a magnetic field.
Finally we consider the effect of the spin orbit interaction
on the phenomenon of  ``reflectionless tunnelling.''

\end{abstract}

\
\bigskip

\section{Introduction}

\noindent
In mesoscopic samples of disordered normal metals at low temperatures
it is possible
to observe a quantum correction
to the classical Boltzmann conductance \cite{Bergmann1}.
Here mesoscopic means that within the sample, the interaction of 
a single electron with other degrees of freedom, such as
other electrons, phonons, magnetic impurities etc., can be neglected.
Such a situation can be realised in semiconductor and metal
nanostructures cooled to milli-Kelvin temperatures.
The origin of the quantum correction
is quantum 
interference between time reversed scattering paths.
(By ``time reversed path'' we mean that the electron follows the
same trajectory but
in the opposite sense.)
The correction is sensitive to the breaking of spin rotation 
symmetry and is 
suppressed by the breaking of time reversal
symmetry \cite{Bergmann2}.
The spin orbit interaction has the important property of breaking 
spin rotation symmetry while preserving time reversal symmetry.

As an example, the two probe conductance   $G=(e^2/h)g$ of a
quasi-one dimensional mesoscopic 
wire  
 can be expressed in 
the form $g= g^{cl} + \delta g$ \cite{Slevin2}.
The classical conductance of the wire is $O(N)$:
  $g^{cl}=N/(1+s)$ where $N$ is the number of scattering 
channels and $s=L/l$ is
the ratio of the length $L$ of the wire to the elastic mean free
path $l$.
In the absence of an applied magnetic field the quantum correction
$\delta g$, also called the ``weak localisation
correction'',
is $O(N^0)$. 
More exactly, in the diffusive regime $1\ll s\ll N$
we have $\delta g = -2/3$ if the spin orbit interaction
is negligible and $\delta g =+1/3$ if it is not. 
In a magnetic field sufficient  to break time reversal
symmetry the quantum correction is 
suppressed and becomes $O(1/N)$. 
 
More recently, a quantum correction to the classical conductance of a
disordered mesoscopic normal superconducting (NS) junction was observed 
\cite{Kastalsky}.
The correction is most pronounced in junctions where the
transparency $\Gamma$ of the Schottky barrier at the NS
interface is low $\Gamma \ll 1$ and the length of the normal 
part is such that $\Gamma s \simeq 1$.
Under these conditions a phenomenon known as ``reflectionless
tunnelling'' occurs.
In \cite{Wees} it is suggested that quantum interference is
responsible for this effect. 
The type of scattering paths involved 
in this quantum interference
contain a segment where the path of an electron 
(hole) incident on the NS boundary is subsequently retraced 
by an Andreev reflected hole (electron.)
The enhancement of the junction conductance can be as much
as several times greater than the classical conductance. 

In this paper we investigate the effect of the coupling of the 
spin and spatial degrees of freedom of the electrons by an appreciable 
spin orbit interaction has on this quantum correction.
As previously mentioned, this has an important effect on the
quantum correction in a normal metal.
As the electron diffuses through the metal, the spin orbit coupling
causes a simultaneous diffusion of the direction of its spin.
Interference between different spin states is now possible and
this modifies the interference responsible for the weak 
localisation correction.
What role then, if any, does this spin diffusion 
play in the NS junction?

At first sight the answer to this question would be appear to be
none at all.
The argument is as follows: the spin rotations of an electron and
an Andreev reflected hole, which traverse time reversed paths,  
are exactly opposite and cancel each other.
Since interference between scattering processes involving
such paths is believed
to be responsible for the quantum correction, it should 
be insensitive to the spin orbit interaction.
As we shall demonstrate below this is too simple and in fact
the spin orbit interaction does affect the quantum correction to
the classical conductance of an NS junction.
We have found the flaw in the argument is that it is 
not only processes involving
electrons and holes traversing time reversed paths which are 
responsible for the quantum correction.
Processes in which electrons and holes traverse loops in the
same sense, what we shall call here identical paths,
also contribute and in this case the spin 
rotations do not cancel each other. 
The most striking consequence of the existence of interference
involving such paths is that the quantum correction is sensitive
to the spin orbit interaction even in a magnetic field.

In the normal metal we have seen that the quantum correction
to the classical conductance is suppressed when time reversal
symmetry is broken in an applied magnetic field.
This is easily understood since the symmetry between time 
reversed paths is broken when the time reversal symmetry is broken.
In the NS junction however, the
contribution of identical paths is still present even in
a magnetic field; the 
electron and hole carry opposite charges, and so the 
Aharonov-Bohm phases 
that they accumulate as they follow such trajectories
cancel.
Moreover, since the electron and hole undergo identical 
spin rotations, as they follow identical paths, this
contribution is sensitive to the spin diffusion induced by the spin
orbit interaction.
 
The paper is organised as follows. In Sections \ref{BdG}
-\ref{formulae}
we deal
with some necessary preliminaries: the Bogolubov-de Gennes 
equations, scattering theory 
and conductance formulae for the NS junction,  
generalising the standard theory 
 to take into account the spin orbit interaction.

In Section \ref{quasiball} we investigate a quasi-ballistic
NS junction,
 that is to say a junction
whose length is shorter than the mean free path but long enough 
so that its scattering matrix is well described by a certain random
phase approximation \cite {rpa}. 
We express the junction conductance $G_{NS}$ as a sum of two 
terms
\begin{equation}
G_{NS} = \frac{2e^2}{h} g_{NS} =
\frac{2e^2}{h} ( g_{NS}^{cl} + \delta g_{NS} )
\end{equation}
a classical conductance $g_{NS}^{cl}$ and a quantum correction
$\delta g_{NS}$ due to quantum interference. 
We determine the quantum correction for the four ensembles listed
in Table \ref{table1}
and the results are presented in Table \ref{table2}.
In Section \ref{interp} we present a semiclassical
interpretation of these results which permits us to identify
the type of scattering paths which interfere to produce
the quantum correction. 
In agreement with Ref \cite{Wees}, we find that there is 
an important contribution
due to interference between processes involving electrons and holes
traversing time reversed paths, 
but that there is also a second contribution from processes involving
electrons and holes traversing identical paths.
We explain why this contribution is sensitive to the breaking of 
spin rotation symmetry but not to the breaking of time reversal
symmetry.

After reaching a firm understanding of the quasi-ballistic junction
we consider the diffusive regime, looking first at junctions without
a Schottky barrier at the NS interface in Section \ref{diffjunc}.
In zero magnetic field the quantum correction
is known to be of $O(N^0)$ \cite{Takane2,Beenakker4,Beenakker2}.
In a magnetic field the authors of Ref \cite{Brouwer} find that
the correction though smaller is still $O(N^0)$.
The effect of the spin orbit interaction was not considered 
in Ref \cite{Brouwer} and so we repeat their calculation 
taking it into account.
We find that breaking spin rotation symmetry multiplies the 
correction by a factor of minus one half,
regardless of whether time reversal symmetry is broken or not.

In Section \ref{reftun} we discuss the dramatic enhancement of the junction conductance, known as ``reflectionless tunnelling,''
which is observable when $\Gamma \ll 1$
and $\Gamma s \simeq 1$.
Under these conditions the contribution of time reversed paths 
is $O(N)$ and dominates that of identical paths which is
$O(N^0)$
so that the reflectionless
tunnelling effect is, in a first approximation,
insensitive to the 
spin orbit interaction.
This conclusion has been confirmed by carrying out a numerical
simulation of a junction under the relevant conditions.

\section{Bogolubov-de Gennes equations and the spin
orbit interaction}
\label{BdG}

\noindent
The Bogolubov-de Gennes (BdG) equations \cite{deGennes}
appropriate for a metal where the electrons' spatial
and spin degrees of freedom are coupled by a significant 
spin-orbit interaction, are
\begin{equation}
\left[
\begin{array}{cc}
H_e, &  \Delta \\ \Delta^* , & {\cal T} H_e {\cal T} 
\end{array} \right]
\left[
\begin{array}{c} \psi^{e}  \\ \psi^{h}  \end{array} \right]
= \epsilon \left[ \begin{array}{c} \psi^{e}  \\ \psi^{h}  \end{array} \right]
\end{equation}
where
\[
H_e=H_0-E_F + \Omega
\]
Here $H_0\equiv H_0(\vec r, \sigma,\mu)$
is the single electron Hamiltonian of the metal
incorporating the spin-orbit interaction and $E_F$ is the Fermi
energy.
In deriving this equation, it is assumed that an attractive point like 
interaction exists between the electrons:
\begin{equation}
\label{BCSV}
V(\vec r_1, \vec r_2) = -\frac{1}{2} V_0 \delta(\vec r_1 - \vec r_2)
\end{equation}
The full interacting Hamiltonian is reduced to an effective
non-interacting Hamiltonian by introducing effective potentials
\begin{equation}
\label{Delta}
\Delta(\vec r) = V_0 \overline{ \Psi(\vec r, \uparrow) 
\Psi(\vec r, \downarrow) }
\end{equation}
\begin{equation}
\label{Omega}
\Omega(\vec r,\sigma,\mu) = V_0 (1-2\delta_{\sigma,\mu})
\overline{ \Psi(\vec r, \sigma) 
\Psi^{\dag}(\vec r, \mu) }
\end{equation}
where the overline indicates a thermal average with respect to the Fermi
distribution function and $\Psi$ is the usual field operator
appearing in the second quantised formulation of the interacting
electron problem.
The time reversal operator ${\cal T}$, which appears in the BdG
equations has the form
\begin{equation}
\label{T}
{\cal T} = \rho C 
\end{equation}
\begin{equation}
\label{rho}
\rho=i\sigma_y = 
\left( \begin{array}{cc} 0 & +1 \\ -1 & 0 \end{array} \right)
\end{equation}
with $C$ the operation of complex conjugation.
The eigenstates of the BdG equation describe the excitations 
of the interacting electron system.
The meaning of the electron $\psi^e$ and $\psi^h$ wavefunctions
can be seen by writing the field operator as
\begin{equation}
\label{Bog.Val.transf.}
\Psi(\vec r, \sigma) = \sum_n  \left[ \psi^{e} _n(\vec r, \sigma) \gamma_n -
{\cal T}[ \psi^{h} _n (\vec r, \sigma) ] \gamma_n^{\dag} \right]  
\end{equation}
where $\gamma_n$ is the annihilation operator of the excitation
labelled $n$.
The corresponding eigenvalue $\epsilon_n$ of the BdG equations
corresponds to the energy of the excitation.
With the aid of (\ref{Bog.Val.transf.}) the occupation 
of the eigenstates of $H_e$ for a given excitation  can be determined.

\section{Scattering theory for the NS junction}
\label{scatter}

\noindent
In this section we develop the scattering theory appropriate to
the normal superconducting junction shown schematically 
in Figure 1.
We treat the scattering at the NS interface within 
Andreev's approximation.
We suppose that any impurities in the system are in the 
region indicated by shading in Figure 1, and
we make the approximation that the magnetic field is zero everywhere
except in this disordered region.
This is reasonable for the low fields of interest which affect 
the interference
between electrons and holes in the normal part of the junction.

\subsection{Scattering matrices for electrons and holes}

\noindent
To facilitate the explanation of the formalism, 
it is helpful to develop the scattering theory for a 
definite model.
We have chosen a lattice tight binding model, the same model
which we will use later in numerical simulations.
An exact analogous explanation is also possible for a continuum model.

We consider a cubic lattice and take into account nearest
neighbour interactions only.
We denote by $\psi^{e} (x,y,z,\sigma)$ the amplitude that the 
electron is in an s- orbital at $(x,y,z)$ with spin
$\sigma$ and similarly $\psi^{h} (x,y,z,\sigma)$
for a hole. 
We include in the Hamiltonian a spin orbit term which arises from the 
Zeeman coupling of the electron spin with the effective magnetic field 
felt by the electron as it moves in the spatially varying potential of 
the lattice.
We ignore the direct Zeeman coupling of the spin to any external 
magnetic field.
A simple calculation shows that the Hamiltonian has the form: 
\begin{equation}
\begin{array}{llll}
<x,y,z,\sigma| H_e | x,y,z,\mu> & = & & E_0 \delta_{\sigma,\mu} \\
<x,y,z,\sigma| H_e | x-1,y,z,\mu> & = & & v^x_{\sigma,\mu} \\
<x,y,z,\sigma| H_e | x,y-1,z,\mu> & = & &v^y_{\sigma,\mu} \\
<x,y,z,\sigma| H_e | x,y,z-1,\mu> & = & \exp(-i \alpha x) & v^z_{\sigma,\mu}
\end{array}
\label{model}
\end{equation}
where
\begin{equation}
\begin{array}{lll}
 v^x_{\sigma,\mu} & = V_0 \delta_{\sigma,\mu} -  V_1  i [\sigma_x]_{\sigma,\mu} \\
 v^y_{\sigma,\mu} & = V_0 \delta_{\sigma,\mu} -  V_1  i [\sigma_y]_{\sigma,\mu} \\
 v^z_{\sigma,\mu} & = V_0 \delta_{\sigma,\mu} -  V_1  i [\sigma_z]_{\sigma,\mu}
\end{array}
\end{equation}
An external magnetic field $B$, applied in the $+y$ direction 
is modelled by Peierl's factors in the matrix elements between nearest
neighbours in the $z$ direction.
If $a$ is lattice constant $\alpha=2 \pi Ba^2/\phi_0$ where 
$\phi_0=h/e$ is the flux quantum. 
The transverse dimensions are $1\le x \le L_x$ and $1\le y \le L_y$.
The Hamiltonian (\ref{model}) can be regarded as a three dimensional
generalisation of that proposed in Ref. \cite{Ando} as a model
for a two dimensional electron gas formed at the surface
of a III-V semiconductor.

The relevant energy scale of the model is determined by $V_0^2+V_1^2$.
For convenience we shall set this to unity with the choice
\begin{equation}
\begin{array}{cc} V_0=\cos \theta & V_1 =\sin\theta \end{array}
\end{equation}
Varying the angle $\theta$, we may set an arbitrary ratio of normal
potential coupling $V_0$ to spin orbit coupling $V_1$, while
keeping the extent in energy of the density of states roughly constant.
With this choice, $v^x$,$v^y$ and $v^z$ are all elements of $SU(2)$.
Using the homomorphism of $SU(2)$ with the
three dimensional rotation group $SO(3)$, we can interpret the
$v$'s as rotations of the spin of the electron as it moves
between nearest neighbours \cite{Arfken}.
A product of nearest neighbour matrix elements along a
path will have the form $\exp(i \Phi) v$
where $\Phi$ is the Aharonov Bohm phase picked up by the electron,
and $v\in SU(2)$ is the 
rotation of the electron's spin as it traverses the path.

The Hamiltonian $H_e$ has the form given in (\ref{model}) everywhere
except in the disordered region located in $0<z<L$.
There, some or all of the Hamiltonian matrix elements are supposed
random. 
In Ref. \cite{Ando} the diagonal elements were assumed to
be independently and identically distributed, while the
parameter $\theta$ controlling the spin orbit interaction
was held fixed.
In Refs. \cite{Zanon,Evangelou} a random spin orbit interaction
was also considered.
For the present purpose we do not need to specify
the precise distribution.

 First, we consider the scattering of electrons incident at an energy 
$E=E_F+\epsilon$.
To the left of the disordered section we expand the electron 
wavefunction $\psi^{e}$ in terms of the Bloch states of (\ref{model}) with 
energy $E$. 
\begin{equation}
\psi^{e}(x,y,z,\sigma) = \sum_{n;\text{Im}k_n\le0} 
a_{+n} \psi_{+n}(x,y,\sigma) \exp (+i k_n z)
+
\sum_{n;\text{Im}k_n=0}
a_{-n} 
\psi_{-n}(x,y,\sigma) \exp (-i k_n z)
\end{equation}
\begin{equation}
\psi_{-n}(x,y,\sigma) = \sum_{\mu} \rho_{\sigma,\mu} \psi_{+n}(L_x-x+1,y,\mu) 
\label{chi L}
\end{equation}
As $z\rightarrow - \infty$, far from the disordered region, we 
impose the boundary condition that the allowed states consist 
exclusively of incoming and outgoing propagating waves.
Thus in $z<0$ states with $\text{Im}k_n>0$ are excluded.  
We denote by $2N$ the number of  ``open channels", i.e. states with 
$\text{Im}k_n=0$  that
carry a positive probability current in the $+z$ direction;
there are an equal number carrying current in the $-z$ direction.
We label these states so that $+n$ carries a current in the $+z$ direction 
and $-n$ a current in the $-z$ direction. 
After a suitable normalisation of
the transverse wavefunctions (see Appendix \ref{A}) 
the electric current due to electrons at 
the left of the disordered section is 
\begin{equation}
\label{Ie}
I_e = \frac{-e}{h} \sum_{n;\text{Im}k_n=0} |a_{+n}|^2 - |a_{-n}|^2
\end{equation}
The boundary condition as $z\rightarrow -\infty$, imposed above,
ensures that only open channels, and not ``closed channels'' with
$\text{Im}k_n\neq0$, contribute to the current.
A similar expansion may be made on the right in terms of
a set of coefficients $\{a_{+n}^{\prime}, a_{-n}^{\prime}\}$ with
the boundary condition that we admit only those states with
$\text{Im}k_n\ge0$. Thus, far to the right of the disordered 
section, as $z\rightarrow +\infty$, the allowed states again consist
exclusively of incoming and outgoing propagating waves.

The $4N\times 4N$ scattering matrix for electrons $S_e$ relates the 
$4N$ incoming
flux amplitudes at the left, $a_+=\{a_{+n};\text{Im}k_n=0\}$ and
the right $a_-^{\prime}=\{a_{-n}^{\prime};\text{Im}k_n=0\}$ 
with the $4N$ outgoing flux amplitudes at the left
$a_-=\{a_{-n};\text{Im}k_n=0\}$ and the right
$a_+^{\prime}=\{a_{+n}^{\prime};\text{Im}k_n=0\}$

\begin{equation}
\label{defn Se}
S_e
\left[ \begin{array}{c} a_+ \\ a_-^{\prime} \end{array} \right]
=
\left[ \begin{array}{c} a_- \\ a_+^{\prime} \end{array} \right]
\end{equation}
The matrix $S_e$ has the structure
\begin{equation}
\label{Se}
S_e = \left[ \begin{array} {cc}
r_e  &  t_e ^{\prime} \\
t_e  &  r_e ^{\prime} \end{array} \right] 
\end{equation}
in terms of the $2N \times 2N$reflection and transmission matrices for 
left incidence
($r_e , t_e$) and right incidence ($r_e ^{\prime}, t_e ^{\prime}$).

Since we are considering time independent scattering,
the currents to the left and right of the disordered section
must be equal, and therefore it follows that $S_e$ is unitary.
There is an additional restriction on $S_e$ when, 
in the absence of an applied magnetic field,
the Hamiltonian
is time reversal invariant ie. $[H_e,{\cal T}] =0$ 
with ${\cal T}$ given in Eq. (\ref{T}).
For a suitable choice of transverse wavefunctions (see Appendix \ref{A})
$S_E$ will then satisfy
\begin{equation}
\label{S vs ST}
S_e^T = 
\left[ \begin{array}{cc} 
1_{2N}  &  0  \\
0   & -1_{2N} 
\end{array}\right]
S_e
\left[\begin{array}{cc} 
-1_{2N}  &  0  \\
 0   &  1_{2N} 
\end{array}\right] 
\end{equation}
where $1_{2N}$ means the $2N\times 2N$ unit matrix.
This can be written in the equivalent form 
\begin{equation}
\label{r vs rT}
\begin{array}{lll}
r_e                    & =-  & r_e^T \\
r_e^{\prime}           & =-  &(r_e^{\prime})^T \\
t_e                    & =+  &(t_e^{\prime})^T \\
\end{array}
\end{equation}
The simplicity of these relations, compared with
for example those of Ref \cite{Mello2}, is related to the 
presence of $\rho$ in Eq.
(\ref{chi L}) (see Appendices \ref{A} and \ref{B})

We now turn to the scattering matrix $S_h$ for the holes.
From the BdG equations we can see that the hole wavefunction
$\psi^{h} $ evolves according to 
\begin{equation}
H_h \psi^{h}  = i \hbar \frac{\partial}{\partial t} \psi^{h}  
\end{equation}
where $H_h$ is given by
\begin{equation}
H_h[+B] = {\cal T} H_e[+B] {\cal T} = -H_e[-B] 
\end{equation}
If $\psi^{h} $ describes the scattering of a hole with excitation energy
$+\epsilon$ in a field $+B$ then
${\cal T}\psi^{h} $ describes the scattering of an electron
at energy $-\epsilon$ also in field $+B$.
We shall make use of this in two ways.

Firstly, outside the disordered region $B=0$ and $H_h=-H_e$. Thus outside
the disordered region it is useful to expand $\psi^{h} $ in terms of the 
Bloch states of $H_e$ at energy $E_F-\epsilon$. 
At the left, for example
\begin{equation}
\psi^{h} (x,y,z,\sigma) = \sum_{n;\text{Im}k_n\le0} 
b_{-n} \psi_{+n}(x,y,\sigma) \exp (+i k_n z)+\sum_{n;\text{Im}k_n=0}
b_{+n} \psi_{-n}(x,y,\sigma) \exp (-i k_n z) 
\end{equation}
Note that since in this region $H_h=-H_e$ the probability currents are 
reversed by comparison with the electron case. 
We therefore associate the coefficient $b_{+n}$, the flux amplitude for a 
positive hole probability current in the $+z$ direction, with the wavefunction
proportional to $\exp (-i k_n z)$.
The holes carry an opposite electric charge to that of the electrons
so that they carry an electric current
\begin{equation}
\label{Ih}
I_h = \frac{e}{h} \sum_{n;\text{Im}k_n=0} | b_{+n}|^2 - |b_{-n}|^2  
\end{equation}
By definition, the $4N \times 4N$ matrix $S_h$ relates incoming hole 
probability currents to outgoing probability currents
\begin{equation}
\label{defn Sh}
S_h
\left[ \begin{array}{c} b_+ \\ b_-^{\prime} \end{array} \right]
=
\left[ \begin{array}{c} b_- \\ b_+^{\prime} \end{array} \right] 
\end{equation}
The matrix $S_h$ has a structure similar to that of Eq. (\ref{Se}); i.e.
\begin{equation}
\label{Sh}
S_h =  \left[ \begin{array} {cc}
r_h  &  t_h ^{\prime} \\
t_h  &  r_h ^{\prime} \end{array} \right] 
\end{equation}

Secondly by rewriting ${\cal T}\psi^{h} $ in terms of electron flux amplitudes, and 
recalling that these amplitudes are related by $S_e$, we arrive at a 
relation between $S_e(-\epsilon, +B)$ and $S_h(+\epsilon, +B)$
\begin{equation}
S_h(+\epsilon, +B) = 
\left[
\begin{array}{cc} 1_{2N} & 0 \\ 0 & -1_{2N} \end{array}
\right]
S_e^*(-\epsilon, +B) 
\left[
\begin{array}{cc} -1_{2N} & 0 \\ 0 & 1_{2N} \end{array}
\right] 
\end{equation}
or
\begin{equation}
\label{Se vs Sh}
\begin{array}{lll}
r_h(+\epsilon,+B)             & =- & [r_e(-\epsilon,+B)]^* \\
r_h^{\prime}(+\epsilon,+B)    & =- & [r_e^{\prime}(-\epsilon,+B)]^* \\
t_h(+\epsilon,+B)             & =+ & [t_e(-\epsilon,+B)]^* \\
t_h^{\prime}(+\epsilon,+B)    & =+ & [t_e^{\prime}(-\epsilon,+B)]^*
\end{array} 
\end{equation} 
Again we assume here a suitable choice of transverse wavefunctions
(see Appendix \ref{A}.)
Following a similar line of argument it is possible to demonstrate a relationship 
between $S_e(+\epsilon,+B)$ and $S_e(+\epsilon,-B)$.

\subsection{Andreev scattering at the NS interface}

\noindent
In this section we outline the calculation of the coefficients
of Andreev reflection
\cite{Andreev} at the NS interface.
These are essentially unchanged by the introduction of a 
spin orbit interaction in the materials forming the junction.
We assume throughout that $\Delta_0\ll E_F$, a condition
which is realised in practice.

Far from the NS interface in the normal metal the superconducting gap 
$\Delta\rightarrow 0$.
On the other hand, far from the NS interface in the superconductor, 
$\Delta\rightarrow \Delta_0 \exp(i \varphi)$
where $\Delta_0$ is real.
In general the reflection coefficients will depend on the precise 
form of $\Delta$ in the transition region near the junction.
For a point contact junction, however, it is permissible to 
assume a simple step model

\begin{equation}
\label{Delta step model}
\begin{array}{ll}
\Delta=0 & z \notin \text{S} \\
\Delta=\Delta_0 \exp(i \varphi) & z \in \text{S}
\end{array}
\end{equation}
We are interested in the situation where the energy 
$E=E_F+\epsilon$ of the incident 
electron
is in the energy gap of the superconductor $E_F<E<E_F+\Delta_0$.
Anticipating somewhat in order to avoid unnecessary algebra, we find 
the electron is mainly reflected as a hole like excitation.
A solution of the BdG equation in the normal metal corresponding to
this is
\begin{equation}
\label{Andreev in normal metal}
\left[ \begin{array}{c} \psi^{e}  \\ \psi^{h}  \end{array} \right]
=
\left[ \begin{array}{c} 1 \\ 0 \end{array} \right]
\exp (i k_n^{(+)} z) \psi_{+n}^{(+)}(x,y,\sigma) +  
\left[ \begin{array}{c} 0 \\ r_{he}^A \end{array} \right]
\exp (i k_n^{(-)} z) \psi_{+n}^{(-)}(x,y,\sigma) 
\end{equation}
where $r_{he}^A$ denotes the matrix of Andreev reflection amplitudes 
and the superscript $(\pm)$ refer to Bloch states with energies
$E^{(\pm)} = E_F \pm \epsilon$.
The first term describes an excitation where an electron above
the Fermi level is incident from the left in channel $n$.
The second term corresponds to an excitation in which an electron below
the Fermi level  
is annihilated, i.e. to a reflected ``hole'' with opposite velocity
and spin to that of the incoming electron.
Since $\Delta_0 \ll E_F$ we can to a good approximation ignore
the difference between $k_n^{(+)}$ and $k_n^{(-)}$ and
similarly the differences between the transverse wavefunctions.
Requiring that the wavefunction and its derivative be
continuous at the boundary of the superconductor leads to
\begin{equation}
\label{rA he 2}
\begin{array}{ll}
r_{he}^A = i\exp(-i \varphi) \; , & \; \; \; r_{eh}^A = i\exp(+i \varphi) 
\end{array}
\end{equation}
Here we have made the further assumption that $\epsilon\ll \Delta_0$,
the limit of interest in what follows, and we have also
given the reflection coefficient for an incident hole.

\subsection{The Scottky barrier at the NS interface}

\noindent
In real NS junctions, a mismatch between the conduction bands of the two
materials which make up the junction results in 
the creation of a Schottky barrier at the interface.
This barrier plays an important role in the physics of
the device and so we must take it into account.
We shall model the Schottky barrier as a simple planar potential
barrier.
At the barrier an incident particle may be either transmitted
without a change of momentum or specularly reflected.
We neglect any dependence of the reflection and transmission
probabilities on the momentum of the incident
particle so that the properties of the barrier are described 
by a single parameter $\Gamma \in [0,1]$,
its transparency.
The transmission and reflection matrices which make up
the electron scattering matrix $S^{B}_e$ of the barrier
are
\begin{equation}
\label{r,t B}
\begin{array}{lccl}
t_e^{B}          & = & \sqrt{\Gamma}  &1_{2N} \\
t_e^{\prime B}   & = & \sqrt{\Gamma}  &1_{2N} \\
r_e^{B}          & = & \sqrt{1-\Gamma}& w_B \\
r_e^{\prime B}   & = & -\sqrt{1-\Gamma}& w_B^{\dagger}
\end{array}
\end{equation}
The precise form of the $2N \times 2N$ matrix $w_B$ depends
on the choice of transverse wavefunctions.
For subsequent analysis we need only note, however, that $w_B$
is antisymmetric and unitary. 
The hole scattering matrix $S^{B}_h$ for the barrier is
related to $S^{B}_e$ in the usual way by (\ref{Se vs Sh}).

\subsection{Combination of scattering matrices}
\label{scomb}

\noindent
It is the purpose of this section, having considered above the scattering
matrices for the various components of the NS junction, to explain how 
the scattering matrices may be combined to find the total 
scattering matrix. 
We consider first a junction without a Schottky barrier.
For the normal part we can write
\begin{equation}
\label{defn S eh}
\left[ \begin{array} {cc}
r &  t ^{\prime} \\
t  &  r ^{\prime} \end{array} \right] 
\left[ \begin{array}{c}  c_+ \\  c_- ^{\prime} \end{array} \right]
=
\left[ \begin{array}{c}  c_- \\  c_+ ^{\prime} \end{array} \right] 
\end{equation}
where 
\begin{equation}
\label{r eh}
r = \left[ \begin{array} {cc}
r_e  &   0  \\
0     &  r_h  \end{array} \right]   \;\; etc.
\end{equation}
are $4N$-dimensional matrices and 
\begin{equation}
\label{c}
\begin{array}{ccc}
c_{\pm} 
=  \left[ \begin{array}{c} a_{\pm } \\ b_{\pm} \end{array} \right] \; ,&
c_{\pm}^{\prime}   
=  \left[ \begin{array}{c} a_{\pm }^{\prime} \\ b_{\pm}^{\prime} \end{array} \right]
\end{array} 
\end{equation}
$4N$-dimensional vectors, in the notation of Sec. IV.
For the Andreev part we define the 
$4N\times 4N$ scattering matrix $S^A$ by
\begin{equation}
\label{defn SA}
S^A  c_+ ^{\prime}= c_- ^{\prime}
\end{equation}
By reference to Section III. this has the form
 \begin{equation}
\label{SA}
S^A = 
\left[ \begin{array} {cc}
0 & r^A_{eh} 1_{2N} \\
r^A_{he} 1_{2N} & 0
\end{array} \right] 
\end{equation}
For the combined system, the $4N \times 4N$  $S$ matrix is defined by
\begin{equation}
\label{defn S}
S  c_+ = c_-    
\end{equation}
In equation (\ref{defn S eh}) we replace $c_- ^{\prime}$ in terms of 
$c_+ ^{\prime}$ as in (\ref{defn SA}); we then perform the matrix 
multiplication and, from the two resulting equations, eliminate 
$ c_+^{\prime}$. From the expression relating $ c_+$ and 
$c_-$ we extract 
the $4N \times 4N$ $S$ matrix of (\ref{defn S}) to obtain
\begin{equation}
\label{combined S}
S= r + t^{\prime} S^A \frac{1}{1-r^{\prime}S^A}t
\end{equation}
In the electron-hole spaces, $S$ 
of  Eq. (\ref{combined S}) has the structure
\begin{equation}
\label{S}
S = \left[ \begin{array} {cc}
r_{ee}  &  r_{eh} \\
r_{he}  &  r_{hh} \end{array} \right] 
\end{equation}
$r_{ee}$ etc. being $2N \times 2N$ matrices.
To determine the conductance we shall need the submatrix $r_{he}$.
From Eqs. (\ref{combined S}) and (\ref{S}) we find 
\begin{equation}
\label{rhe 1}
r_{he}=t_h ^{\prime}\left(S^A \frac{1}{1-r^{\prime} S^A}
\right)_{he} t_e 
\end{equation}
Using the structure (\ref{SA}) of $S^A$ we can write $r_{he}$ as
\begin{equation}
\label{rhe 2}
r_{he}=t_h^{\prime}r_{he}^A 
\left(\frac{1}{1- r^{\prime} S^A}\right)_{ee} t_e  
\end{equation}
For any nonsingular operator $D$ we now use the operator identity 
\cite{feshbach}
\begin{equation}
\label{identity1}
\left( \frac{1}{D} \right) _{ee}= \left( \frac{1}
{D_{ee}-D_{eh}\frac{1}{D_{hh}}D_{he}} \right)_{ee}  
\end{equation}
to find
\begin{equation}
\label{rhe 3}
r_{he} = t_h^{\prime} r_{he}^A \left[ 1 - r_e^{\prime} r_{he}^{A} 
r_h^{\prime} r_{eh}^A \right]^{-1} t_e
\end{equation}

For a junction with a Schottky barrier we first consider the  
composition of the Schottky barrier and the superconductor.
The scattering matrix for this system can be obtained from
Eq. (\ref{combined S}) where $r$, $t$ and 
$t^{\prime}$
are taken from the model (\ref{r,t B}) for the barrier. 
With the aid of the identity (\ref{identity1}) and the 
following one \cite{feshbach}
\begin{equation}
\label{identity2}
\left( \frac{1}{D}\right) _{he}= -\frac{1}{D_{hh}}D_{he}
\left( \frac{1}{D_{ee}-D_{eh}\frac{1}{D_{hh}}D_{he}}\right)_{ee}
\end{equation}
we find
\begin{equation}
S^{BS} = \left[ \begin{array} {cc}
r_{ee}^{BS}  &  r_{eh}^{BS} \\
r_{he}^{BS}  &  r_{hh}^{BS} \end{array} \right] 
\end{equation}
\begin{equation}
\label{r BS}
\begin{array}{l}
r_{he}^{BS} = i \exp(-i \varphi) \left[ \Gamma/(2-\Gamma) \right] 1_{2N}   \\
r_{eh}^{BS} = -(r_{he}^{[BS]})^*  \\
r_{ee}^{BS} = 2 \left[ \sqrt{1-\Gamma}/(2-\Gamma) \right] w_B   \\
r_{hh}^{BS} = -(r_{ee}^{[BS]})^*
\end{array}
\end{equation}
The scattering matrix for the complete system 
of normal part, Schottky
barrier and superconductor can now be obtained 
from  Eq. (\ref{combined S}),
where $r^{\prime}$, etc. refer to the normal metal as before, 
but $S^A$ is 
replaced by $S^{BS}$ of (\ref{r BS}).
(Note that in deriving Eq. (\ref{combined S}) the structure 
(\ref{SA}) of $S^A$ was not used.)

\section{Conductance formulae}
\label{formulae}

\noindent
We assume that the bias voltage is small in comparison
to $\Delta_0/e$, and that the size of the superconducting part 
is long enough so that there is no quasi-particle current in the 
superconductor. In this case, the zero temperature dc conductance 
$G_{NS}$ of the normal-superconducting junction can be described 
by a simplified ``Landauer'' formula which has been derived in 
Ref.\cite{BTK,Lambert,Takane1} 
\begin{equation}
G_{NS} = \frac{2 e^2}{h}g_{NS} =
\frac{2 e^2}{h} \text{tr} r_{he} r_{he}^{\dag} 
\label{gns1}
\end{equation}
where $r_{he}$ is the $2N \times 2N$ matrix of electron-hole reflection amplitudes for the 
composite system. 

There is an important simplification if the 
Hamiltonian of the system is time reversal invariant and the 
bias voltage is small in comparison to the Thouless energy $E_c$
\cite{thouless},
so that the energy dependence of $S_e$ can be neglected.
The conductance then has the form \cite{Beenakker1}
\begin{equation}
\label{gns2}
G_{NS} = 
\frac{2 e^2}{h} \sum_{n=1}^{2N} \left[ \frac{T_n}{2-T_n} \right]^2 
\end{equation}
where $T_n$ are the eigenvalues of $t_e t_e^{\dag}$.
We have verified that this result still holds when spin 
rotation symmetry is broken by the spin orbit
interaction.

In what follows we wish to compare the quantum conductance of
the NS junction, calculated from (\ref{gns1}), with the classical
conductance of the junction.
This latter quantity is determined using the classical rule of 
combining conductances 
\begin{equation}
1/g = 1/g_1 + 1/g_2
\end{equation}
This corresponds to the addition of flux intensities as opposed to
flux amplitudes.
The conductance associated with the electron traversing the normal
part is $2N/s$ and similarly for the hole in the traversing the
normal part in the opposite direction.
The conductance of the barrier is $2N |r^{BS}_{he}|^2$, so that the classical
conductance is
\begin{equation}
\label{gcl}
g_{NS}^{cl}=\frac{2N|r^{BS}_{he}|^2}{1+2s|r^{BS}_{he}|^2} 
\end{equation}
The classical conductance is insensitive to the breaking of time reversal and
spin rotation symmetries.

\section{Quasi-ballistic junction}
\label{quasiball}

\noindent
In this section we shall calculate the conductance of an NS junction 
to first order in $s=L/l$, where $L$ is the length of the
normal part 
of the junction and $l$ is the elastic mean free path. 
We neglect terms of order $s^2$ 
and above so that result is strictly applicable only in the limit that
$s\ll 1$.
Nevertheless we shall see that the results of the calculation shed
considerable
light on the origin of the quantum interference in the device.

The $2N\times 2N$ reflection matrix $r_{he}$ for the system consisting 
of the normal metal, barrier and superconductor can be obtained as
discussed in Section \ref{scomb}.
After expanding to second order in $r_e ^{\prime}$,  $r_h ^{\prime}$,
which is sufficient for an evaluation of $g_{NS}$ to first order
in $s$, 
we have
\begin{equation}
\begin{array}{lll}

r_{he} &  =  & t^{\prime}_h r_{he}^{BS} t_e 

+ t^{\prime}_h r_{he}^{BS} r_e^{\prime} r_{ee}^{BS} t_e

+ t^{\prime}_h r_{hh}^{BS} r_h^{\prime} r_{he}^{BS} t_e

+ t^{\prime}_h r_{he}^{BS} r_e^{\prime} r_{ee}^{BS}  r_e^{\prime}
  r_{ee}^{BS}  t_e \\

& + & t^{\prime}_h r_{hh}^{BS} r_h^{\prime} r_{he}^{BS}  r_e^{\prime}
  r_{ee}^{BS} t_e 

+ t^{\prime}_h r_{hh}^{BS} r_h^{\prime} r_{hh}^{BS}  r_h^{\prime}
  r_{he}^{BS}  t_e

+ t^{\prime}_h r_{he}^{BS} r_e^{\prime} r_{eh}^{BS}  r_h^{\prime}
  r_{he}^{BS}  t_e

+ \dots
\end{array}
\label{mss}
\end{equation} 
The conductance is found by substituting this into (\ref{gns1}) and
performing an average over an ensemble of scattering matrices $S_e$,
describing a set of microscopically different but macroscopically
equivalent configurations of impurities in the normal 
part of the junction.

In principle, the distribution for $S_e$ should be calculated
from some model distribution of Hamiltonians. We shall not, however, 
 attempt to do that here.
Instead we will assume that the resulting ensemble of scattering 
matrices $S_e$ is distributed according to the ``local maximum
entropy model'' \cite{Slevin2,Mello1,Pichard1}.
If the geometry is quasi-$1d$ and the number of channels $N$ 
sufficiently large, then results obtained with the aid of the local 
maximum entropy model are known to be identical
to those obtained from the class of 
microscopic models
described by the the nonlinear sigma model \cite{Frahm,Frahmpete}.

The distribution of $S_e$ in the local model depends on $N$, $s$
and the symmetry of the Hamiltonian, i.e. whether or not time reversal 
symmetry is broken and whether or not spin rotation symmetry is broken.
There are {\it four} ensembles (Table. \ref{table1}.)
The critical strengths of the magnetic field and the spin orbit 
interaction separating the various ensembles should be similar to
those associated with the weak localisation effect in normal metals.
The details of the distribution of $S_e$ for the four ensembles
can be found in Appendix \ref{C}.

It will be helpful, when we come to discuss the  
interpretation of the results, to write (\ref{mss}) in
the form
\begin{equation}
r_{he}= \sum_{i=1}^{\infty} p_i
\end{equation}
Each term in the series represents the contribution of a
particular scattering processes to $r_{he}$.
The classical conductance is obtained by ignoring interference
between different processes and summing intensities
\begin{equation}
g_{NS}^{cl} =  \hbox{tr} \left< \sum_{i=1}^{\infty} p_i p_i^{\dagger}
\right>
\end{equation}
The quantum correction to this classical conductance is found
by summing the interference between different processes
\begin{equation}
\delta g_{NS} = \hbox{tr} \left<  \sum_{i\neq j} p_i p_j^{\dagger} \right>
\label{mss2}
\end{equation}
After carrying out the average we find that
\begin{equation}
g^{cl}_{NS} = 2N |r^{BS}_{he}|^2 (1 - 2 |r^{BS}_{he}|^2 s + O(s^2) )
\end{equation}
which agrees with the expansion of (\ref{gcl}) to the order
we are considering. 
The explicit expressions for $\delta g_{NS}$ are collected together
in Table \ref{table2}.
The function $f$ of the barrier transparency, which appears in
the table, has the explicit form: 
\begin{equation}
f(\Gamma) = \frac{2 \Gamma^2}{(2-\Gamma)^2} 
\left[
\frac{2 \Gamma^2}{(2-\Gamma)^2} -1 
\right]
\end{equation}
There are two obvious limiting cases: $\Gamma=1$ corresponding to 
a junction without a Schottky barrier and $\Gamma\ll 1$ corresponding
to a junction with a high Schottky barrier.
The first point to note is that the quantum correction is of $O(N)$, 
the same order as the classical conductance, in zero field.
Secondly for $\Gamma\ll 1$, the conductance increases as disorder 
is added to the junction.
This is the essence of the dramatic reflectionless tunnelling 
effect which we discuss in Section \ref{reftun}.
Thirdly when time reversal symmetry
is broken by the application of a magnetic field the quantum correction
is $O(N^0)$ and not $O(1/N)$ as might have been expected by analogy
with the weak localisation effect in a normal metal.
The final, and perhaps the most surprising result, is that in a magnetic
field the breaking of
spin rotation symmetry by the spin orbit interaction multiplies the
quantum correction by a factor of minus one half
even though the symmetry of the Hamiltonian, in the sense of
random matrix theory, is unchanged and remains unitary.

\section{Semiclassical Interpretation}
\label{interp}

\noindent
The importance of quantum interference between processes
in which the path of an electron 
(hole) incident on the NS boundary is subsequently retraced 
by an Andreev reflected hole (electron)
was first pointed out in \cite{Wees}. 
In the absence of a magnetic field, and if the bias voltage is small enough, 
electrons which move along a path in one given sense are phase
conjugated with holes traversing the time reversed path.
In a magnetic field, or if the bias voltage is large enough,
this phase conjugation is destroyed.
However we have seen that there is a significant quantum correction 
even in a magnetic field.
There must therefore be an additional source of quantum interference 
which is not sensitive to the breaking
of time reversal symmetry.
As we shall see the relevant processes involve paths
in which an electron (hole) and an Andreev reflected
hole (electron) traverse a loop in the same sense.
In order to remain concise, we shall refer to such processes as
containing identical paths.
The interference involving such paths can only be destroyed 
by applying a large enough bias voltage.
The physical importance of the bias voltage as a ``symmetry
breaking parameter'' is discussed in \cite{Brouwer}.

Only some of the terms in (\ref{mss2}) are found to be nonzero
after averaging, so that in fact
\begin{equation}
\delta g_{NS} =  \hbox{tr} \left< p_1 p_5^{\dagger} + p_1 p_7^{\dagger} +
p_5p_1^{\dagger} + p_7 p_1^{\dagger} \right> + O(s^2)
\end{equation}
where $p_j$ means the $j$th term in (\ref{mss}.)
Consider first the interference
between process $p_1$ and $p_5$.
An example of a scattering path which contributes to process $p_1$
\begin{equation}
p_1 = t^{\prime}_h  r_{he}^{BS}  t_e
\end{equation}
is illustrated in Figure 2. Since we are working to
first order in $s$ it is sufficient to consider the motion of the electron
and holes, traversing from one side of the normal part of the 
junction to the other, as ballistic, so these trajectories appear 
as straight lines in Figure 2.
Examples of paths which contribute to $p_5$ 
\begin{equation}
p_5 = t^{\prime}_h r_{hh}^{BS} r_h^{\prime} r_{he}^{BS}  r_e^{\prime}
  r_{ee}^{BS}  t_e
\end{equation}
are illustrated in Figures 3 and 4.
To first order in $s$ the interference between processes $p_1$ and
$p_5$ is
\begin{equation}
\hbox{tr} \left< p_1 p_5^{\dagger} +
p_5p_1^{\dagger}   \right>
= 2 | r_{he}^{BS} |^2 | r_{ee}^{BS} |^2 \hbox{tr}\left< 
r_h^{\prime} r_e^{\prime} \right>
\label{p1p5}
\end{equation}
The remaining terms involving $p_1$ and $p_7$ contribute
\begin{equation}
\hbox{tr} \left< p_1 p_7^{\dagger} +
p_7p_1^{\dagger}   \right>
= - 2 | r_{he}^{BS} |^4  \hbox{tr}\left< 
r_h^{\prime} r_e^{\prime} \right>
\end{equation}

In the interest of brevity we will concentrate 
on the interference between $p_1$ and $p_5$.
A very similar analysis is possible for the interference
between processes $p_1$ and $p_7$ leading to identical
conclusions.
We proceed by relating the product of electron and
hole reflection matrices in (\ref{p1p5}) to a product
of an electron and a
hole Greens function.
To simplify the algebra we shall suppose that both the spin orbit 
interaction and the magnetic field are zero everywhere
except in the disordered region.
We shall also suppose that $L_y=1$ and impose periodic boundary
conditions in the $x$ direction.
The Bloch states at energy $E$ have the form
\begin{equation}
\begin{array}{llllll}
\psi_{2m} (x,z,\sigma) &= &\exp(ik^x_{2m} x) \exp(ik_{2m} z) 
\delta_{\sigma,\sigma_{2m}} & \sigma_{2m}&=&\uparrow \\
\psi_{2m+1} (x,z,\sigma) &=& \exp(ik^x_{2m+1} x) \exp(ik_{2m+1} z) \delta_{\sigma,\sigma_{2m+1}} & \sigma_{2m+1}&=&\downarrow \\
\psi_{-2m} (x,z,\sigma) &=& -\exp(-ik^x_{2m} x) \exp(-ik_{2m} z) \delta_{\sigma,\sigma_{-2m}} & \sigma_{-2m} &=&\downarrow \\
\psi_{-(2m+1)} (x,z,\sigma) &=& \exp(-ik^x_{2m+1} x) \exp(-ik_{2m+1} z) \delta_{\sigma,\sigma_{-(2m+1)}} & \sigma_{-(2m+1)}&=&\uparrow 
\end{array}
\label{basis}
\end{equation}
where
\begin{equation}
k^x_{2m} = k^x_{2m+1} = \frac{2\pi}{L_x} \text{   } m=0,\dots,L_x-1
\end{equation}
and the energy and the momenta are related by
\begin{equation}
E= 2 \cos k^x_m + 2 cos k_m
\end{equation}
The reflection matrices for electrons and holes can be related
to the corresponding Green's functions as indicated in Appendix \ref{A}.
\begin{eqnarray}
[r_e^{\prime}]_{mn} & = & -i \sqrt{4\sin k_m \sin k_n} \exp(+i (k_m+  
k_n)L) \nonumber \\
& & \sum_{x_a \sigma x_b \sigma^{\prime}}
\psi_{+m}^*(x_b\sigma) G_e^+(x_b,L,\sigma; x_a,L,\sigma^{\prime})
\psi_{-n} (x_a \sigma^{\prime}) 
\label{Gere}
\end{eqnarray}
\begin{eqnarray}
[r_h^{\prime}]_{nm} & = & -i \sqrt{4\sin k_m \sin k_n} \exp(-i(k_m+  
k_n)L) \nonumber \\
& & \sum_{x_a \sigma x_b \sigma^{\prime}}
\psi_{-n}^*(x_b\sigma) G_h^+(x_b,L,\sigma; x_a,L,\sigma^{\prime})
\psi_{+m} (x_a \sigma^{\prime}) 
\label{Ghrh}
\end{eqnarray}
Note that, for convenience, the states (\ref{basis}) have not been  
normalised to carry identical currents,
proper account of this has been taken in the expressions (\ref{Gere})
and (\ref{Ghrh}) for the reflection matrices.

The quantum correction involves a trace over the product of $r_e$ and $r_h$.
Using the relation with the Green's function this can be separated into two
elements: an integration over the cross section involving the transverse
wavefunctions and an average of the product of an electron and a hole
Greens function.
We will consider the second element first. This involves evaluating
\begin{equation}
\Lambda_{\sigma_{-n},\sigma_{+m}} (x_a,x_b) =
\left<  G_h^+(x_a,L,\sigma_{-n}; x_b,L,\sigma_{+m})
G_e^+(x_b,L,\sigma_{+m}; x_a,L,\sigma_{-n}) \right>
\label{gfavg}
\end{equation}
Within the semiclassical approximation,
as is explained in Ref. \cite{gutz}, Ch. 12
and Ch. 13, we can express the 
Greens functions as summations over paths:
\begin{equation}
\begin{array}{lcll}
G_e^+(x_b,L,\sigma_{+m}; x_a,L,\sigma_{-n}) & = &
\sum_{j:x_a\rightarrow x_b}
A_j \exp(i S_j + i \Phi_j) [v_j]_{\sigma_{+m},\sigma_{-n}} \\
G_h^+(x_b,L,\sigma_{-n}; x_a,L,\sigma_{+m}) & = & 
-\sum_{j:x_a\rightarrow x_b}
A_j \exp(-i S_j - i \Phi_j) [v_j]_{\sigma_{-n},\sigma_{+m}}
\end{array}
\label{paths}
\end{equation}
Substituting (\ref{paths}) into (\ref{gfavg}) and
taking the disorder average we find that only two 
contributions remain: 

\subsection{Time reversed paths}

\noindent
The path of the electron moving between $x_a$ and $x_b$ is
retraced by the Andreev reflected hole as illustrated in Figure 3.
 The electron and hole
charges are opposite but they traverse the path in opposite directions
so the Aharonov-Bohm phase factors they accumulate do not cancel 
each other out. Thus this term is important only when the magnetic 
field is negligible
and the Hamiltonian is time reversal symmetric.

Let us consider first the situation where the spin orbit interaction is
negligible.
In the quasi-$1d$ limit, the average of the product of the
hole and electron Green's functions is independent of $x_a$ and
$x_b$ so that
\begin{equation}
\Lambda_{\sigma_{-n},\sigma_{+m}} (x_a,x_b) =
-\lambda \delta_{\sigma_{-n},\sigma_{+m}}
\end{equation}
After integrating the product of transverse wavefunctions over the
cross section we find
\begin{equation}
\left< [r_h^{\prime}]_{mn} [r_e^{\prime}]_{nm} \right>=
4\sin k_m \sin k_n 
\lambda \delta_{\sigma_{-n},\sigma_{+m}} + \text{I.P.}
\label{trps}
\end{equation}
where I.P. denotes the contribution due to identical paths
which we will consider below.
We determine $\lambda$ by demanding consistency
with the local maximum entropy model.
In this model $S_e$ is distributed isotropically (see Appendix
\ref{C})
so that averages, 
such as that in (\ref{trps}),
are independent of the channel indices $n$ and $m$.
The extent to which the isotropy assumption is reasonable for
microscopic models is discussed in Refs. \cite{Slevin1,Rodolfo2}.
In the present case, to maintain consistency with the results
of Section \ref{quasiball}, we are required 
to make the approximation that
the group velocities appearing in (\ref{trps}) can be replaced
by an average velocity.
This is a reasonable assumption if the energy is such that 
we are far from any subband edges, but is of questionable  
validity for energies close to a subband edge.
Having made this approximation and determined $\lambda$ by
comparison with Section \ref{quasiball} we find
\begin{equation}
<(r_h^{\prime})_{2n+1,2m} (r_e^{\prime})_{2m,2n+1}>=
<(r_h^{\prime})_{2n,2m+1} (r_e^{\prime})_{2m+1,2n}>=\frac{s}{N+1}
\end{equation}

If the spin orbit interaction is appreciable we assume that the spin
direction of an electron or hole traversing a path is completely
randomised. 
The matrix $v$ describing this rotation can then be assumed to be
uniformly distributed, relative to the invariant measure, on the
group $SU(2)$.
The result is
\begin{equation}
<(r_h^{\prime})_{n,m} (r_e^{\prime})_{m,n}>=\frac{s}{2N-1}
\end{equation}
After summation over all channels the contribution of this
term to the quantum correction is essentially independent of the
strength of the spin orbit interaction.
This is because the spin rotation experienced by an electron
traversing a path in a given sense cancels that of the hole 
traversing the same path in the opposite sense.
To demonstrate this explicitly it suffices to note that after summation over
spin indices the contribution of time reversed paths is proportional
to tr$v^{\dagger}v=\text{tr}1_2$, which is a constant independent
of the distribution of $v$. 
 
\subsection{Identical paths $x_a=x_b$}

\noindent
The electron path in the normal metal contains a loop which
is retraced in the same sense by the Andreev reflected hole
as illustrated in Figure 4.
The Aharonov-Bohm phase accumulated by the electron and hole, are
now exactly opposite and cancel. Therefore this term is important 
even if time reversal symmetry is broken.

Proceeding in a similar way to that above we find that the
contribution of identical paths is
\begin{equation}
\left<  [r_h^{\prime}]_{nm} [r_e^{\prime}]_{mn} \right>=
4\sin k_m \sin k_n 
\lambda \delta_{m,n}
\left< v_{\sigma_{-n},\sigma_{+m}} v_{\sigma_{+m},\sigma_{-n}} \right>
 + \text{T.R.P.}
\label{ips}
\end{equation}
(Here T.R.P. refers to the contribution from time reversed paths.)
The presence of the delta function, arising from the integration of
the transverse functions over the cross section, means 
that identical paths only contribute for electrons
and holes which are back scattered by the disordered region
as is indicated in Figure 4.
Note the novel feature that this is true even if time reversal
symmetry is broken in the disordered region.
The average over the matrices $v$ is calculated with respect to the 
distribution $p(v)=\delta(v-1_2)$ if the spin orbit interaction is
negligible, and $p(v)$ uniform on $SU(2)$, if it is appreciable.
As above we determine $\lambda$ by reference to Section \ref{quasiball}.
If the spin orbit interaction is negligible we find: 
\begin{equation}
<(r_h^{\prime})_{2n+1,2m}(r_e^{\prime})_{2m,2n+1}>=
<(r_h^{\prime})_{2n,2m+1}(r_e^{\prime})_{2m+1,2n}>=
\delta_{m,n} \frac{s}{N+1}
\end{equation}
in zero magnetic field and: 
\begin{equation}
<(r_h^{\prime})_{2n+1,2m}(r_e^{\prime})_{2m,2n+1}>=
<(r_h^{\prime})_{2n,2m+1}(r_e^{\prime})_{2m+1,2n}>=
\delta_{m,n} \frac{s}{N}
\end{equation}
in a nonzero field.
If the spin orbit interaction is appreciable we find
\begin{equation}
<(r_h^{\prime})_{n,m}(r_e^{\prime})_{m,n}>=
\delta_{m,n} \frac{-s}{2N-1}
\end{equation}
in zero magnetic field and
\begin{equation}
<(r_h^{\prime})_{n,m}(r_e^{\prime})_{m,n}>=
\delta_{m,n} \frac{-s}{2N}
\end{equation}
in a nonzero field.
The change of sign when spin rotation symmetry is broken comes about
because an electron and a hole traversing the
same path in the same direction undergo the same spin rotation.
After summation over the spin indices, the contribution from 
identical paths is thus proportional to tr$v^2$ which is
sensitive to the distribution of $v$
\begin{equation}
\begin{array}{llll}
\text{tr}v^2 & = & +2 &  p(v)=\delta(v-1_2)           \\
             & = & -1 &  p(v)\text{ uniform on } SU(2)
\end{array}
\end{equation}

\section{Diffusive junction without a Schottky barrier $(\Gamma=1)$}
\label{diffjunc}

\noindent
For convenience we will defer discussion of a junction for which the Scottky 
barrier has a low transparency ($\Gamma\ll 1$) and which exhibit 
reflectionless tunnelling, to Section VIII, and focus in this section on
junctions without a Schottky barrier.

We now have to sum the perturbation series (\ref{mss})
to all orders in $s$ to find the quantum 
correction for a diffusive junction.
In the absence of a magnetic field, this amounts to the evaluation
of the integral: 
\begin{equation}
g_{NS} = \int_{0}^{1} \text{d} T \rho (T)  \left[ \frac{T}{2-T} \right]^2
\label{gfromden}
\end{equation}
Here $\rho(T)$ is the density of the eigenvalues of the matrix $t_e t_e^{\dagger}$,
where $t_e$ refers to the composite system composed of 
the normal metal and Schottky barrier. 

Within the local maximum entropy
 model, the development of the density $\rho(T)$ 
as a function of $s$ is
described by a nonlinear diffusion equation \cite{mello pichard}.
The initial condition on $\rho(T)$
at $s=0$ is: 
\begin{equation}
\rho_{s=0}(T) = N \delta (T-1)
\label{initcond}
\end{equation}
In the diffusive regime the solution to this equation is independent
of the ensemble up to corrections of $O(N^0)$ \cite{Beenakker3}.
Evaluating (\ref{gfromden}) we find an identical $O(N)$ 
contribution to $g_{NS}$ for the Orthogonal and Symplectic Ensembles,
while the following $O(N^0)$ term is ensemble dependent.
For the details, we refer the reader to \cite{Beenakker3}. 
With the aid of this reference we obtain the results for the 
Orthogonal and
Symplectic Ensembles given in Table \ref{table3}. 

For a diffusive junction in a magnetic field the summation
of the perturbation series requires the performance
of a number of averages over the unitary group.
The authors of Ref. \cite{Brouwer} find for
the Unitary I Ensemble and $\Gamma=1$ that
\begin{equation}
g_{NS} = \frac{2N}{1+2s} - \frac{8s^3+12s^2+12s}{3(1+2s)^3}
+O(1/N) 
\end{equation}
This gives the same result as Section \ref{quasiball}, in the limit that
$s\ll 1$ and yields $\delta g_{NS} = -1/3$ in the diffusive regime
($1\ll s \ll N$.)
As for the quasi-ballistic junction, the quantum correction 
is $O(N^0)$, not $O(1/N)$, in a magnetic field.

The analysis of Ref \cite{Brouwer}
can be extended to the Unitary II Ensemble.
In the absence of a Schottky barrier we have from (\ref{rhe 3}) and
(\ref{Se vs Sh}): 
\begin{equation}
r_{he} = i \exp(-i\varphi) t^{\prime *} \sum_{p=0}^{\infty}
\left[ -r_e^{\prime} r_e^{\prime *} \right]^p t_e
\end{equation}
Then with the aid of (\ref{param. S}) and (\ref{gns1})
we obtain for the conductance:
\begin{equation}
g_{NS} = \text{tr} \left[
T \sum_{p=0}^{\infty} \left[ -u \sqrt{R} u^* \sqrt{R} \right]^p
u T u^{\dagger}
\sum_{q=0}^{\infty} \left[ - \sqrt{R} u^T \sqrt{R} u^{\dagger}  \right]^q
\right]
\end{equation}
where $u=u_4^*u_2$ is uniformly distributed on $U(2N)$.
Following Ref \cite{Brouwer} the ensemble average is accomplished
in two steps.
First the average on the unitary group and then the average over
the distribution of $T$.
The final result for the Unitary II Ensemble and $\Gamma=1$ is: 
\begin{equation}
g_{NS} = \frac{2N}{1+2s} + \frac{4s^3+6s^2+6s}{3(1+2s)^3}
+O(1/N) 
\end{equation}
which reproduces the result of Section \ref{quasiball} for $s\ll 1$
and yields $\delta g_{NS}=+1/6$ in the diffusive regime.

The most interesting result is that, like the quasi-ballistic
junction, the quantum correction is sensitive to the spin orbit
interaction even if time reversal symmetry is broken.
Referring to Section \ref{interp}, it seems reasonable to assume 
that it is
interference involving identical paths which
is responsible for this sensitivity.

In order to confirm that quantum correction is sensitive to the
spin orbit interaction, even if the junction is a magnetic field
we have a carried out a numerical simulation.
As a model for the junction we use the Hamiltonian of Section
\ref{scatter}.
The Fermi energy $E_F$ is measured
from centre $E_0$
of the energy band of the model.
Since the
zero of energy is arbitrary we can set $E_0=0$ for convenience.
In the normal part of the junction, $0\le z \le L-1$, 
we take the diagonal Hamiltonian elements as:  
\begin{equation}
\begin{array}{llll}
<x,y,z,\sigma| H | x,y,z,\mu> & = &  U(x,y,z) \delta_{\sigma,\mu} &
0\le z \le L_z-1 \\
& & & \\
<x,y,z,\sigma| H | x,y,z,\mu> & = &  U_0 \delta_{\sigma,\mu} & z=L
\end{array}
\end{equation}
The potential $U(x,y,z)$ is the random potential due to impurities
and $U_0$ the height of the Schottky barrier.
For simplicity we shall assume a spatially uncorrelated potential
with a distribution $P(U)=p(U)\text{d}U$
\begin{equation}
\begin{array}{lll}
p(U) & = 1/W & -W/2 \le U \le +W/2 \\
     &       &                     \\
p(U) & = 0 ,    & \text{otherwise.}
\end{array}
\end{equation}
To the right, in $z\ge L+1$, is the superconductor.
The conductance is calculated using the standard Green's function
iteration technique to produce $S_e$ and then using the formulae
of Sections \ref{scatter} and \ref{formulae}. 

In Figure 5 we present the data for the junction
conductance versus the parameter $\theta$,
 defined in Section
\ref{scatter}, controlling
the spin orbit interaction.
The transverse dimensions of the junction are $L_x=L_y=5$ and 
the longitudinal dimension $L=25$. 
The Fermi energy $E_F=0$.
For these parameters $N=17$.
We are interested in the limit $\Gamma=1$ so we set
$U_0=0$.
A magnetic field corresponding to a flux of $1/25\phi_0$
per lattice
cell is applied perpendicular to the junction
corresponding to five flux quanta through the device.
The distribution of the random potential has width $W=3$ corresponding
to an estimated mean free path $l\simeq 4.3$.
The junction is in the diffusive ($L_z\gg l$) quasi-$1d$ 
($L_x,L_y\leq l$, $L_x,L_y\ll L_z$) regime.
The spin rotations between lattice sites are not random but 
differ depending
on the direction of the displacement.
Since these rotations
do not commute the spin direction of the electron is 
effectively randomised and so, for $\theta$ large enough,
the Symplectic Ensemble should be appropriate.

As the magnitude of the spin orbit interaction is increased 
the junction conductance $G_{NS}$ increases, 
while the normal conductance $G_{N}$ 
remains constant.
This is exactly the behaviour predicted.
The value $\delta G_{NS} \simeq 0.8 e^2/h$ 
obtained in the simulation 
is also in reasonable agreement with theoretical 
value $\delta G_{NS}=e^2/h$.

\section{Reflectionless tunnelling}
\label{reftun}

\noindent
This is an effect observed in diffusive junctions when $\Gamma \ll 1$.
In the absence of a magnetic field we may calculate the conductance
in the same way as for the diffusive junction with $\Gamma=1$
but with the initial condition:
\begin{equation}
\rho_{s=0}(T) = N \delta (T-\Gamma)
\end{equation}
instead of (\ref{initcond}).
At $O(N)$ the result is the same for both the Orthogonal and Symplectic
Ensembles:
\begin{equation}
\begin{array}{ll}
g_{NS} = N\Gamma^2/2 + Ns\Gamma^2 & \text{    } 1\ll s\ll N\text{;}\Gamma s\ll 1 \\
g_{NS} = N/(s+1/\Gamma) & 1\ll s \ll N \text{;} \Gamma s\gg 1
\end{array}
\end{equation}
We expect ensemble dependent corrections to this result at $O(N^0)$,
though we have not calculated them explicitly.
This quantum conductance has to be compared to the classical conductance
obtained from Eq. (\ref{gcl}) by setting $|r_{he}|^2 \simeq \Gamma ^2 /4$: 
\begin{equation}
\label{gcl gamma<1}
g^{cl}_{NS} = \frac{N}{s+2/\Gamma^2}
\end{equation}
For both the Orthogonal and Symplectic Ensembles the conductance
is enhanced above the classical value, with a correction $O(N)$.
Note in particular that in the absence of a field the conductance 
increases as disorder is added to the junction.
When the length of the junction is such that 
$\Gamma s \simeq 1$, the junction conductance reaches its maximum
value of roughly $g_{NS} \simeq N \Gamma/2$. 
The classical value at this length is
$g_{NS}^{cl}\simeq N\Gamma ^2 /2 $.
This very large enhancement of the conductance is
referred to as ``reflectionless tunnelling.''

In practice it is the sensitivity of the reflectionless tunnelling 
effect to a magnetic field which is observed in experiments
\cite{Kastalsky}.
We have seen explicitly in the analysis of the quasi-ballistic
junction that the phase conjugation of electrons and holes 
traversing time reversed 
paths is destroyed in a magnetic field, and only
the $O(N^0)$ correction, due to interference involving
identical paths, should remain.
Thus, by applying a magnetic field, the conductance of the junction 
should be dramatically reduced to its classical value,
within a correction of $O(N)^0$.
This argument also indicates that if we break spin rotation
symmetry, the qualitative behaviour in a 
magnetic field should be unchanged,
and there should only be quantitative corrections at 
$O(N^0)$ due to the spin orbit interaction.  
To confirm this we have carried out a numerical simulation
of a junction under the relevant conditions.

The model used has been described in Section \ref{diffjunc}.
The parameters used in the simulation are $L_x=L_y=6$, $L=48$
and $E_F=0$ for which $N=24$ at $\theta=0$ and $N=22$ at $\theta=
\pi/8$.
We have set $W=2$ and estimate that the mean free path $l\simeq 12$.
The junction is in the diffusive ($L_z\gg l$) quasi-$1d$ 
($L_x,L_y\leq l$, $L_x,L_y\ll L_z$) regime.
In Figure 6 we present the data for the conductance versus magnetic
field obtained in the simulations for the two values of 
spin orbit interaction parameter $\theta$ indicated above.
The magnetic field is given in terms of the flux $\phi$ penetrating
the normal part of the junction.
At z=L there is a potential barrier of height $U_0=3$.
The barrier transparency is a function of $U_0$ and
also has a weak dependence on the 
the spin orbit parameter $\theta$ and the magnetic
flux $\phi$.
The average value of $\Gamma=0.23$ with a variation of
about $ 10\% $ over the parameter range used.
At zero field and zero bias the conductance should be approximately: 
\begin{equation}
G_{NS} \simeq \frac{2e^2}{h} \frac{N \Gamma}{2}\approx 5 \frac{e^2}{h}
\end{equation}
which is in reasonable agreement with the numerical data.
On applying a magnetic field we expect to approach the classical value: 
\begin{equation}
G^{cl}_{NS} = \frac{2e^2}{h} \frac{N}{s+2/\Gamma^2} \approx \frac{e^2}{h}
\end{equation}
This is again in reasonable agreement with the numerical data.
A strong reduction of the conductance is observed both when
spin rotation symmetry is unbroken $\theta=0$ and when it
is broken $\theta=\pi/8$ by the spin orbit interaction.
The critical field corresponds to a flux $\phi\simeq2\phi_0$
penetrating the sample, in agreement with critical field expected
for the suppression of the weak localisation effect in normal metals.
The data confirm that, apart from possible corrections of 
$O(N^0)$, the reflectionless
tunnelling effect is not sensitive to the spin orbit interaction.

\section{Conclusions}
\label{discuss}

We have studied the effect of the breaking of spin rotation
symmetry, as a result of an appreciable spin orbit interaction,
in determining the quantum correction to the classical conductance
of a disordered mesoscopic normal superconducting junction.

The most striking result we have obtained concerns the NS junction 
in an applied magnetic field.
We have found that even if time reversal symmetry is broken, 
the quantum correction is sensitive to the spin orbit interaction.
A semiclassical analysis of a quasi-ballistic junction permitted the
scattering processes, which interfere to produce this effect, 
to be identified. 
They involve scattering paths in which an electron and a
hole traverse a loop in the normal part of the junction
in the same sense, that which we have called here ``identical paths''.
 
\section{Acknowledgements}

Keith Slevin would like to thank the European Commission for 
financial support under the Human Capital and Mobility Programme.
Pier Mello would like to thank the Wissenschaftskolleg zu Berlin 
for its hospitality, as well as the Service de Physique de 
l'\'Etat Condens\'e, CEA-Saclay, for the financial support given 
during his visit to Saclay in the early stages of this research. We 
would also like to thank Piet Brouwer for his helpful
comments.
\appendix

\section{}
\label{A}

\noindent
In this appendix we group together certain technical details concerning the scattering
theory for model (\ref{model}) which appears in the main text.
We adopt the Dirac notation so that, for example, $<xyz\sigma|\psi> = \psi(xyz\sigma)$.
The particle current $J_p(z)$ through a cross section at $z$ can be
shown to be:  
\begin{equation}
J_p(z) = \frac{1}{i \hbar} < \psi | D[z] | \psi>
\end{equation}
where $D$ is the operator:
\begin{eqnarray}
D[z] & = & \sum_{xy\sigma\mu} |xyz\sigma><xyz\sigma|H_e|xyz+1\mu><xyz+1\mu|
\nonumber \\
& & -|xyz+1\sigma><xyz+1\sigma|H_e|xyz\mu><xyz\mu|
\end{eqnarray}
By rewriting the Schroedinger equation in the contact in terms of
a real space transfer matrix, and considering the orthogonality relations
between left and right eigenvectors of this matrix, it is possible to
show that the Bloch states satisfy the orthogonality relation: 
\begin{equation}
<m|D[z]|n> = 0  \text{ if }k_m\neq k_n^*
\label{orthog}
\end{equation}
In addition for real $k$ the transverse wavefunctions may be normalised so
that: 
\begin{equation}
<\pm m|D[z]|\pm n> = \pm i \delta_{n,m}
\label{norm}
\end{equation}
Each channel then carrys a charge current of $e/\hbar$. Here we use 
the notation that: 
\begin{equation}
<xyz\sigma|\pm m> = \psi_{\pm m}(xy\sigma) \exp(\pm i k_m)
\end{equation}

In the absence of a magnetic field the Hamiltonian $H_e$ will commute with the time
reversal operator $\cal{T}$.
In this case if $\psi^e$ is a solution to the scattering problem we may construct another
solution $\cal{T}\psi^e$ by operating on $\psi^{e}$ 
with ${\cal T}$.
Taking note of this we can derive the following condition on $S_e$
if the Hamiltonian is time reversal invariant: 
\begin{equation}
S_e^T = 
\left[
\begin{array}{cc} d^{\dag} & 0 \\ 0 & -d^{\dag} \end{array}
\right]
S_e
\left[
\begin{array}{cc} -d & 0 \\ 0 & d \end{array}
\right]
\end{equation}
where $d$ is a $2N \times 2N$ symmetric unitary matrix with elements: 
\begin{equation}
d_{m,n} = -i \sum_{x,y,\sigma,\mu}
\psi_{+m}^*(L_x-x+1,y,\sigma) 
\left[ [v^{z \dag}]_{\sigma,\mu} \exp(i k_m)
- [v^z]_{\sigma,\mu} \exp(-i k_n) \right]
\psi_{+n}^*(x,y,\mu)
\end{equation}
The condition on the scattering matrix when $H_e$ is time reversal invariant, 
 can be simplified by an appropriate choice of basis.
Since $d$ is symmetric unitary it may be decomposed (not uniquely)
as $d=e^Te$ with $e$ unitary. If we do this and make a transformation 
to the new basis:  
\begin{equation}
\psi^{\prime}_{+n}= \sum_{m} e_{n,m} \psi_{+ m}  
\label{newbasis}
\end{equation}
the matrix $d$ becomes unity $d=1_{2N}$.
This transformation is legitimate since it leaves the  current
normalisation unchanged.
The general relation between the electron and hole scattering matrices 
arrived at is, as indicated in the main text:
\begin{equation}
S_h(+\epsilon,+B) = 
\left[
\begin{array}{cc} d^T & 0 \\ 0 & -d^T \end{array}
\right]
S_e^*(-\epsilon,+B) 
\left[
\begin{array}{cc} -d^* & 0 \\ 0 & d^* \end{array}
\right] 
\end{equation}
which again may be simplified by the transformation (\ref{newbasis}).

We now turn to the relation between the electron scattering matrix 
$S_e$ and the 
electron Green function $G_e^+$.
Let us suppose that $H=H_e+U$ with $U$ a random potential. 
To each incoming state at the left $|+n>$ we associate a scattering state 
$|\phi_{+n}>$:
\begin{equation}
|\phi_{+n}> =  |+n> + G_e^+ U |+n>
\end{equation}
Supposing that the random potential $U$ is nonzero only in the volume $\cal{V}$
between $z_0 <z<z_1$ this can be transformed to:
\begin{eqnarray}
|\phi_{+n}> & = &
G^+_e D[H_e,z_1] |+n> - G^+_e D[H_e,z_0-1] |+n> \nonumber \\
& & + \sum_{xyz\notin \cal{V} \sigma} |xyz\sigma><xyz\sigma|+n>
\end{eqnarray}
In $z\le z_0+1$ the scattering state has the form:
\begin{equation}
|\phi_{+n}> =  |+n> + \sum_{m} [r_e]_{m,n} |-m>
\end{equation}
and in the region $z\ge z_1-1$:
\begin{equation}
|\phi_{+n}> =  \sum_{m} [t_e]_{m,n} |+m>
\end{equation}
With the aid of (\ref{orthog}) and (\ref{norm}) we now find:
\begin{equation}
\begin{array}{lcl}
\left[r_e\right]_{mn} & = & -i <-m| D[z_0] G^+_e D[z_0-1] |+n>    \\
\left[t_e\right]_{mn} & = & +i <+m| D[z_1-1] G^+_e D[z_0-1] |+n>  \\
\left[r^{\prime}_e\right]_{mn} & = & -i <+m| D[z_1-1] G^+_e D[z_1] |-n> \\
\left[t^{\prime}_e\right]_{mn} & = & +i <-m| D[z_0+1] G^+_e D[z_1] |-n> 
\end{array}
\end{equation} 
The last two relations are obtained by considering an
incoming wave $|-n>$ from the right.
In a similar way a relation between $S_h$ and $G_h^+$ can be derived. We note only the expression
for the reflection matrix at the right:
\begin{equation}
[r^{\prime}_h]_{mn} = -i <-m| D[z_1-1] G^+_h D[z_1] |+n>
\end{equation}
and we now assume that the magnetic field is zero except in $\cal{V}$.

\section{}
\label{B}

\noindent
As mentioned in the text, the simplicity of (\ref{Se vs Sh}) is related
to the presence of $\rho$ in Eq.(\ref{chi L}).
In most references dealing with random matrix approaches to conduction
in disordered solids definitions, such as those of ref. \cite{Mello2}, 
$\rho$ does not appear in the equivalent of these equations.
To avoid confusion we describe explicitly the transformation between
the sets of definitions.

In the absence of a magnetic field, the scattering matrix of
Ref. \cite{Mello2} which we shall call  
$S_e ^{MP}$ (containing reflection and transmission matrices $r_e ^{MP}$, 
$t_e ^{MP}$, etc.) has the form: 
\begin{equation}
\label{SMP}
S_e ^{MP} =  \left[ \begin{array} {cc}
r_e ^{MP} &  t_e ^{\prime MP} \\
t_e  ^{MP} &  r_e ^{\prime MP} \end{array} \right] 
\end{equation}
and the property that: 
\begin{equation}
\label{SMP vs SMPT}
(S_e^{MP})^T = 
\left[ \begin{array}{cc} 
Z^T  &  0  \\
0   & Z^T
\end{array}\right]
S_e^{MP}
\left[ \begin{array}{cc} 
Z  &  0  \\
 0   &   Z
\end{array}\right] 
\end{equation}
where the $2N\times 2N$ antisymmetric unitary 
matrix $Z$ satisfies (\ref{z1}) and
(\ref{z2}).
The coefficients $a_+$, $a_+ ^{\prime}$ introduced in Section IV
are identical 
to those of Ref. \cite{Mello2}, while the coefficients  
$a_-$, $a_- ^{\prime}$ are related by:  
\begin{equation}
\label{a-}
\begin{array}{c}
a_- ^{MP} = Z a_- \\
a_- ^{\prime MP} = Z a_- ^{\prime} 
\end{array} 
\end{equation}
The relation between $S$ matrices is then: 
\begin{equation}
\label{S vs SMP}
S_e  = 
\left[ \begin{array}{cc} 
Z^T  &  0  \\
0   &   1
\end{array}\right]
S_e^{MP}
\left[ \begin{array}{cc} 
1  &  0  \\
 0   &   Z
\end{array}\right] 
\end{equation}
so that (\ref{SMP vs SMPT}) and (\ref{S vs SMP}) give the relation 
(\ref{S vs ST}).

In the absence of spin-orbit coupling  the two spin components would be 
independent and the reflection and transmission matrices would be block 
diagonal so that, in a suitable basis, 
$S^{MP}$ would have the structure: 
\begin{equation}
\label{MP SO=0}
S_e ^{MP} =  \left[ \begin{array} {cccc}
\tilde r_e  & 0 &  \tilde t_e^{\prime} & 0  \\
0  &  \tilde r_e  & 0 &  \tilde t_e^{\prime}  \\
\tilde t_e  & 0 &  \tilde r_e ^{\prime} & 0  \\
0  & \tilde t_e  & 0 &  \tilde r_e ^{\prime}   \\
\end{array} \right] 
\end{equation}
Here, $\tilde r_e$, $\tilde t_e$, $\ldots$ are $N\times N$ matrices that 
correspond to the 
reflection and transmission matrices for either spin direction and have the 
usual properties, so that $S_e ^{MP}$ is symmetric. The $S$ of this 
paper, obtained from the transformation (\ref{S vs SMP}), would have 
the structure:  
\begin{equation}
\label{SMP SO=0}
S_e  =  \left[ \begin{array} {cccc}
0  &  - \tilde r_e  &   \tilde t_e{\prime} & 0  \\
\tilde r_e  & 0 & 0  & \tilde t_e{\prime}  \\
\tilde t_e  & 0 & 0 & \tilde r_e ^{\prime}   \\
0  & \tilde t_e  &  - \tilde r_e ^{\prime} & 0  \\
\end{array} \right] 
\end{equation}

\section{}
\label{C}

The parameterisations of $S_e$ appropriate for the various ensembles
are discussed in Refs. \cite{Slevin2,Mello2,Mello1,harold1,rodolfo1}.
For the Unitary II Ensemble $S_e$ may be parameterised:  
\begin{equation}
\label{param. S}
S_e = \left[ 
\begin{array}{cc}
u_1\sqrt{R} u_3 & -u_1 \sqrt{T} u_4 \\
u_2 \sqrt{T} u_3 & u_2 \sqrt{R} u_4 
\end{array} \right]  
\end{equation}
where the $u$'s are $2N \times 2N$ unitary matrices,  
$T=\text{diag}\{T_n;n=1,\dots,2N\}$ and $R=1_{2N}-T$.
(The parameters $T_n$ appearing here are the same as 
the eigenvalues of $t_e t_e^{\dag}$ appearing in the
conductance formulae of Section VI.)
The unitary matrices are independently and uniformly 
distributed with respect to the 
invariant measure on the group $U(2N)$ of $2N\times 2N$ 
unitary matrices. 

In the Symplectic Ensemble, time reversal symmetry imposes further
restrictions on the form of $S_e$ and
the appropriate parameterisation is: 
\begin{equation}
S_e = \left[ 
\begin{array}{cc}
u_1 \sqrt{R} Z u_1^T & -u_1 \sqrt{T} Z u_2^T \\
u_2 \sqrt{T} Z u_1^T & u_2 \sqrt{R} Z u_2^T 
\end{array} \right]  
\end{equation}
with $u_1$ and $u_2$ uniformly distributed on $U(2N)$
and $Z$ is a fixed antisymmetric $2N \times 2N$ matrix satisfying: 
\begin{equation}
[Z,T]=0 
\label{z1}
\end{equation}
\begin{equation}
Z^2=-1_{2N} 
\label{z2}
\end{equation}

If the spin orbit interaction is negligible 
the spin is conserved during the electron motion and
the Hamiltonian and scattering matrices can be block diagonalised.
Thus in the Unitary I Ensemble a basis can be found such that: 
\begin{equation}
S_e = \left[ 
\begin{array}{cccc}
 0 &  -\tilde u_1 \sqrt{\tilde R}  \tilde u_3 &
 \tilde u_1 \sqrt{ \tilde T}  \tilde u_4 & 0   \\
 \tilde u_1 \sqrt{ \tilde R}  \tilde u_3 & 0   &
 0 & \tilde u_1 \sqrt{ \tilde T}  \tilde u_4   \\
 \tilde u_2 \sqrt{ \tilde T} \tilde u_3 & 0 &
0 & -\tilde u_2 \sqrt{ \tilde R} \tilde u_4 \\
 0 & \tilde u_2 \sqrt{ \tilde T} \tilde u_3  &
\tilde u_2 \sqrt{ \tilde R} \tilde u_4 & 0 
\end{array} \right]  
\end{equation}
where $\tilde u_1, \dots , \tilde u_4$ are $N\times N$ unitary
matrices uniformly distributed on the unitary group $U(N)$ and
$\tilde T = \text{diag}\{\tilde T_n ;n=1,\dots,N\}$ and $\tilde R=1_{N}-
\tilde T$. 

Finally in the Orthogonal ensemble $S_e$ is parameterised: 
\begin{equation}
S_e = \left[ 
\begin{array}{cccc}
 0 &  -\tilde u_1 \sqrt{\tilde R}  \tilde u_1^T &
 \tilde u_1 \sqrt{ \tilde T}  \tilde u_2^T & 0   \\
 \tilde u_1 \sqrt{ \tilde R}  \tilde u_1^T & 0   &
 0 & \tilde u_1 \sqrt{ \tilde T}  \tilde u_2^T   \\
 \tilde u_2 \sqrt{ \tilde T} \tilde u_1^T & 0 &
0 & -\tilde u_2 \sqrt{ \tilde R} \tilde u_2^T \\
 0 & \tilde u_2 \sqrt{ \tilde T} \tilde u_1^T  &
\tilde u_2 \sqrt{ \tilde R} \tilde u_2^T & 0 
\end{array} \right]  
\end{equation}
with $\tilde u_1$ and $\tilde u_2$ uniformly distributed on $U(N)$.

The distribution of the $T_n$'s is given by the solution of a
Fokker Planck equation, the precise form of which depends on
the ensemble under consideration.
The joint distribution, and in particular the correlations,
of the $T_n$'s depend on the symmetry of the ensemble under
consideration.
However, all we need to know for the present purpose is that: 
\cite{Mello3,chalker} 
\begin{equation}
\left< \sum_{n=1}^{2N} T_n \right> = 2N(1-s) + O(s^2) 
\end{equation}
independent of the ensemble. The only further information
needed to carry out the required averages is that if $u$ is an $N\times N$
unitary matrix uniformly distributed on $U(N)$ then
\begin{equation}
<u_{i,j} u^*_{i^{\prime},j^{\prime}} > = \frac{1}{N} \delta_{i,i^{\prime}} \delta_{j,j^{\prime}}
\end{equation}

An important property of the local maximum entropy
model is that the distribution of $S_e$ is 
isotropic,
that is to say 
the matrices $u_1, \dots , u_4$ are distributed according to
the invariant measure on the unitary group and independently
of the parameters $T_1, \dots , T_{2N}$.

\newpage

\noindent
Figure captions:
\bigskip

\noindent
Figure 1: A schematic of the NS junction for which the scattering theory
is developed in Section \ref{scatter}.
\bigskip

\noindent
Figure 2: An example of a path which contributes to process 
$p_1$ corresponding to
an electron (solid line) traversing the normal part of the junction
whose path is then
retraced by the Andreev reflected hole (dashed line).
\bigskip

\noindent
Figure 3: An example of a ``time reversed path'' which contributes
to process $p_5$.
The path of the electron moving from $x_a$ to $x_b$ 
is retraced by
the Andreev reflected hole as it moves from $x_b$ to $x_a$.
Interference between this path and that illustrated
in Figure 2 is insensitive to
the spin orbit interaction and is suppressed in a magnetic field.
\bigskip

\noindent
Figure 4: An example of an ``identical path'' which contributes
to the process $p_5$.
The electron moves around a loop and returns to
$x_a$. The Andreev reflected hole traverses the loop in
the same direction. 
Interference between this path and that illustrated
in Figure 2 is sensitive to
the spin orbit interaction but not to a magnetic field.
\bigskip

\noindent
Figure 5: The conductance of a quasi-$1d$ diffusive NS junction
as a function of the spin orbit interaction parameter $\theta$ 
in the presence of an applied magnetic field ( flux of $1/25 \phi_0$ 
per lattice cell). The barrier transparency $\Gamma=1$. 
The NS conductance is sensitive to the breaking of spin rotation
symmetry even though time reversal symmetry is broken, while 
the normal conductance is approximately constant as expected.
\bigskip

\noindent
Figure 6: The conductance of a quasi-$1d$ NS junction
as a function of magnetic flux penetrating
the normal part of the junction for zero $\theta=0$
and strong $\theta=\pi/8$ spin orbit
interaction.
The barrier transparency is low ($\Gamma\simeq 0.23$) and the length of the junction is 
such that $\Gamma s \simeq 1$.
We see that the qualitative features of the reflectionless tunnelling
effect are insensitive
to the spin orbit interaction.

\begin{table}[t]
\caption{The various ensembles for which $G_{NS}$ is calculated.
The abbreviation TRS means time reversal symmetry and SRS spin 
rotation symmetry.}
\label{table1}
\begin{tabular}{|l|l|l|}
Ensemble            & TRS & SRS\\
\hline  
Orthogonal  &        yes & yes \\
Unitary I   &        broken  & yes \\
Unitary II  &        broken  & broken  \\
Symplectic  &        yes & broken  \\
\end{tabular}
\end{table}
\bigskip

\begin{table}[t]
\caption{The quantum correction $\delta G_{NS} = (2e^2/h)\delta g_{NS}$
for the quasi-ballistic NS junction
for the ensembles listed in Table 1.}
\label{table2}
\begin{tabular}{|l|l|l|l|}
Ensemble & $\delta g_{NS}$ & $\Gamma=1$ & $\Gamma\ll 1$ \\
\hline
Orthogonal/Symplectic& $-2Ns f(\Gamma)$ & $-4Ns$ & $+Ns\Gamma^2$  \\  
Unitary I            & $-2s f(\Gamma)$  & $-4s$  & $+s\Gamma^2$   \\
Unitary II           & $+s f(\Gamma)$  & $+2s$  & $-s\Gamma^2/2$ \\
\end{tabular}
\end{table}
\bigskip

\begin{table}[t]
\caption{The quantum correction $\delta G_{NS}=(2e^2/h) \delta g_{NS}$ 
for a diffusive ($1\ll s \ll N$) NS junction without a Schottky barrier $(\Gamma=1)$.} 
\label{table3}
\begin{tabular}{|l|ll|}
Ensemble                  & $\delta g_{NS}$ &  \\
\hline  
Orthogonal            &  $\frac{4}{\pi^2}-1$                            &
$(-0.593)$ \\
Symplectic            &  $-\frac{1}{2}\left( \frac{4}{\pi^2}-1 \right)$ &  $(+0.297)$ \\
Unitary I             &   $-\frac{1}{3}$                                &
$(-0.333)$ \\
Unitary II            &   $+\frac{1}{6}$                               &
$(+0.166)$ \\
\end{tabular}
\end{table}


\begin{references}
\bibitem{rpa} The random phase or isotropy assumption for the scattering
matrix is described in Appendix \ref{C} and in Refs. 
\cite{Slevin2,Slevin1,Rodolfo2}
\bibitem{thouless} For a diffusive system the Thouless energy is
$E_c=\hbar D/L^2$ where $D$ is the diffusion constant and $L$ a typical
dimension of the sample. For a ballistic sample $E_c=\hbar v_F/L$
where $v_F$ is the Fermi velocity.
\end{references}
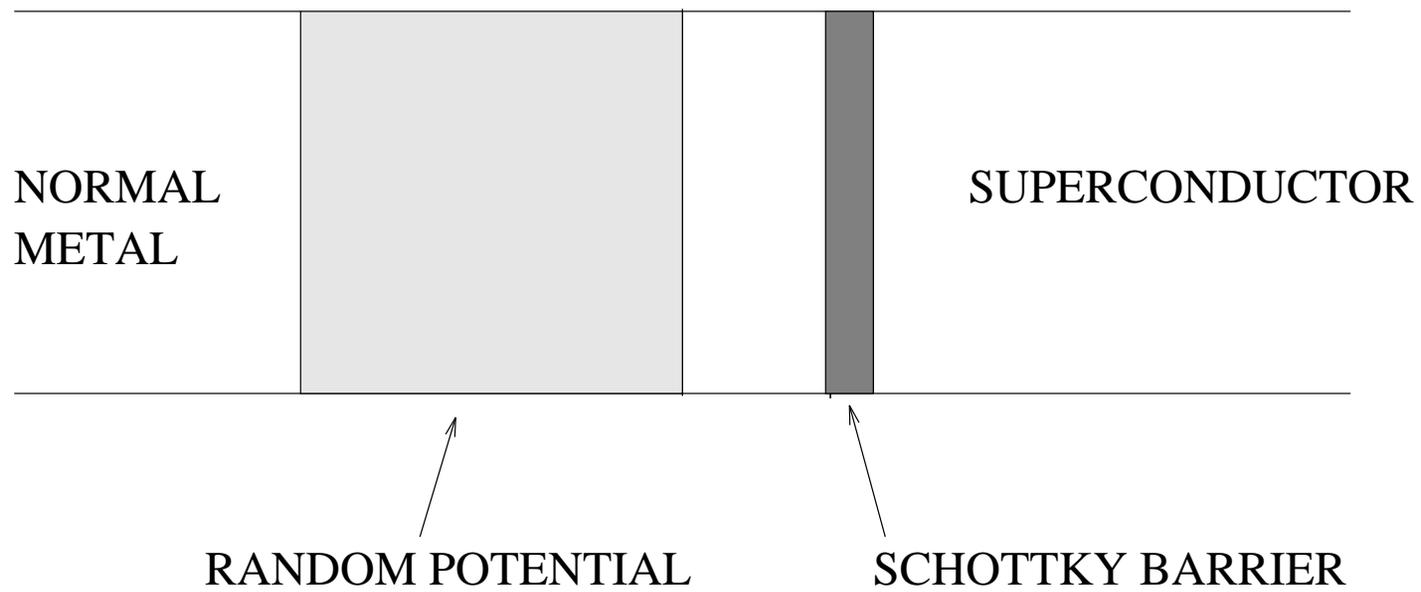
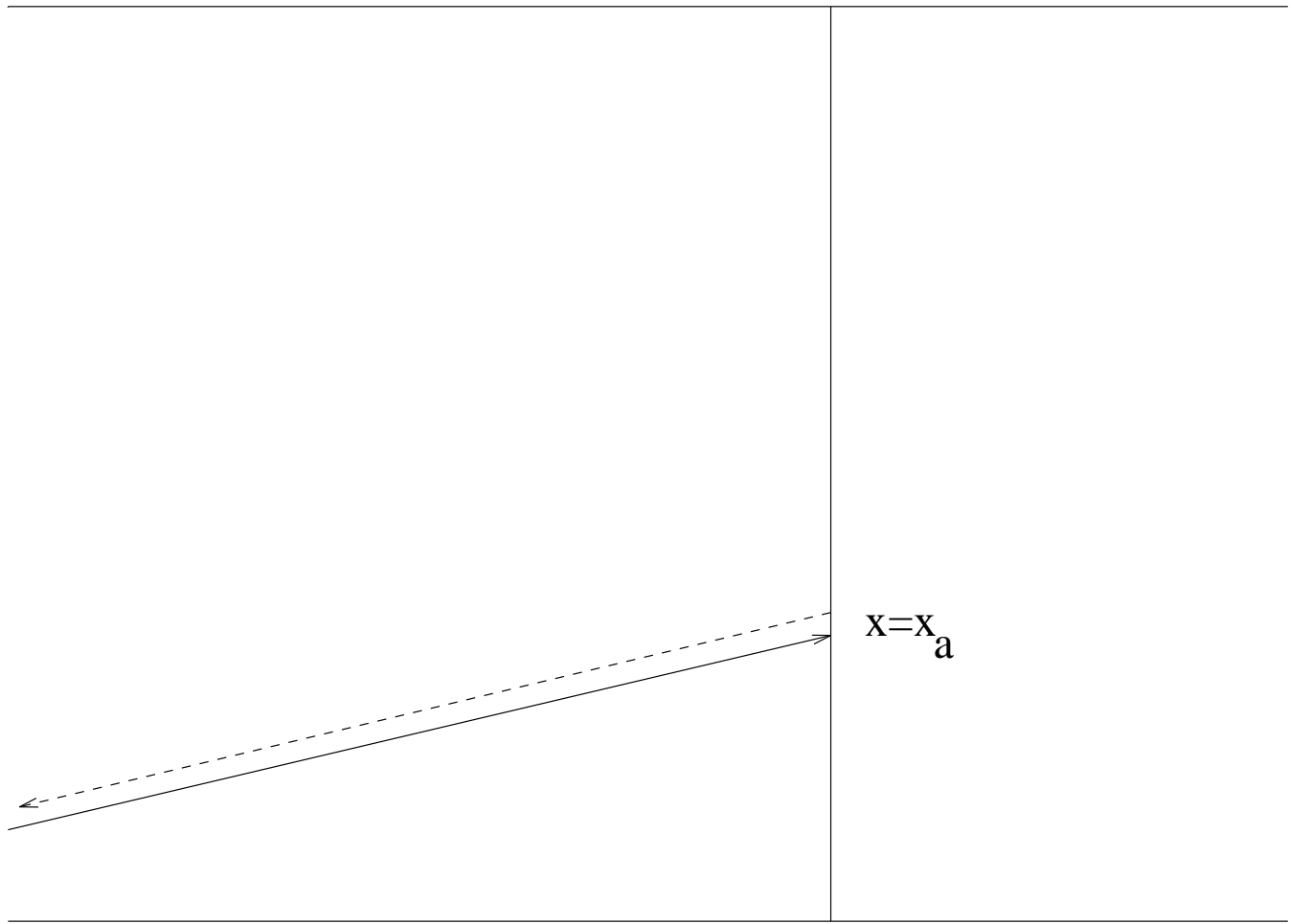
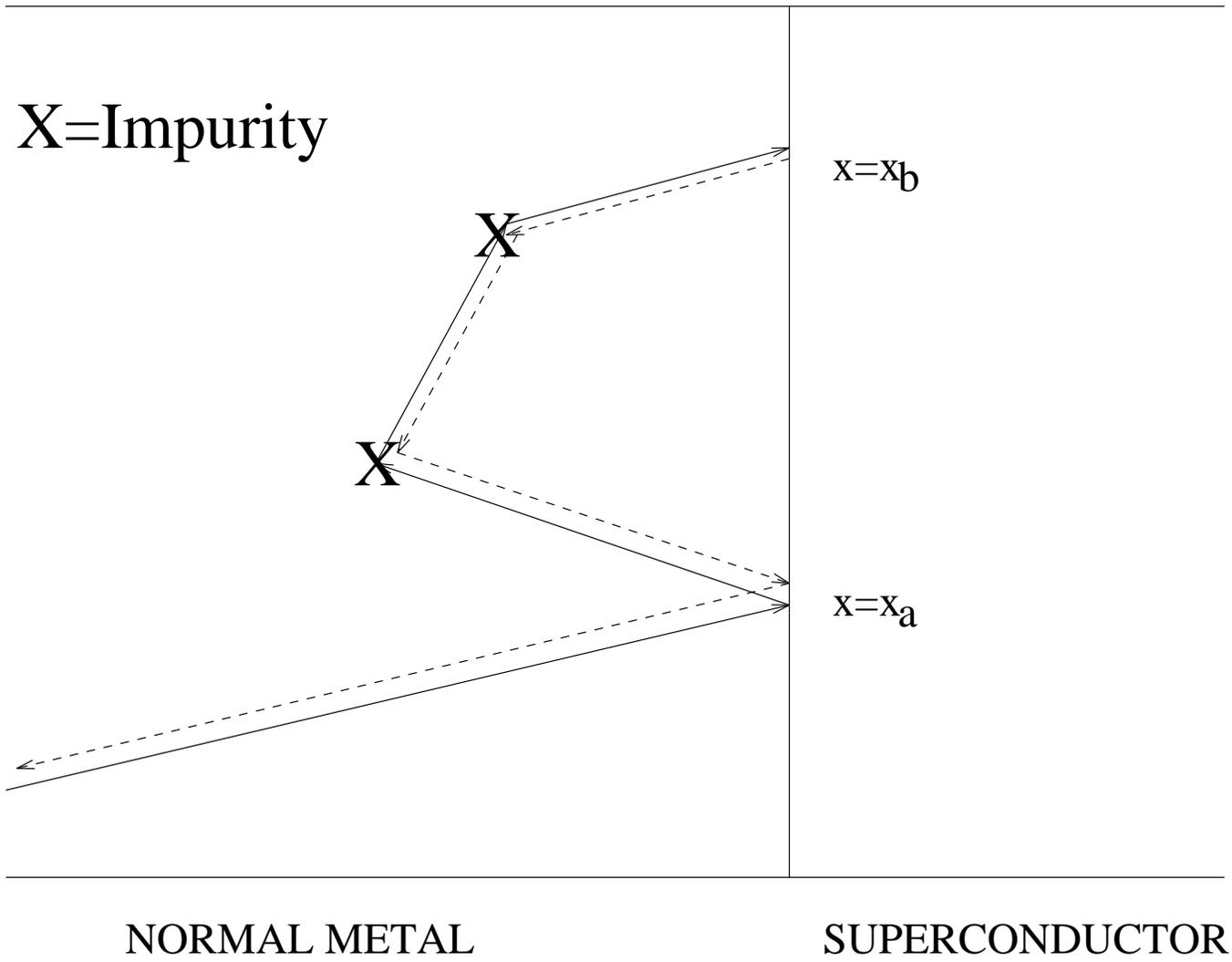
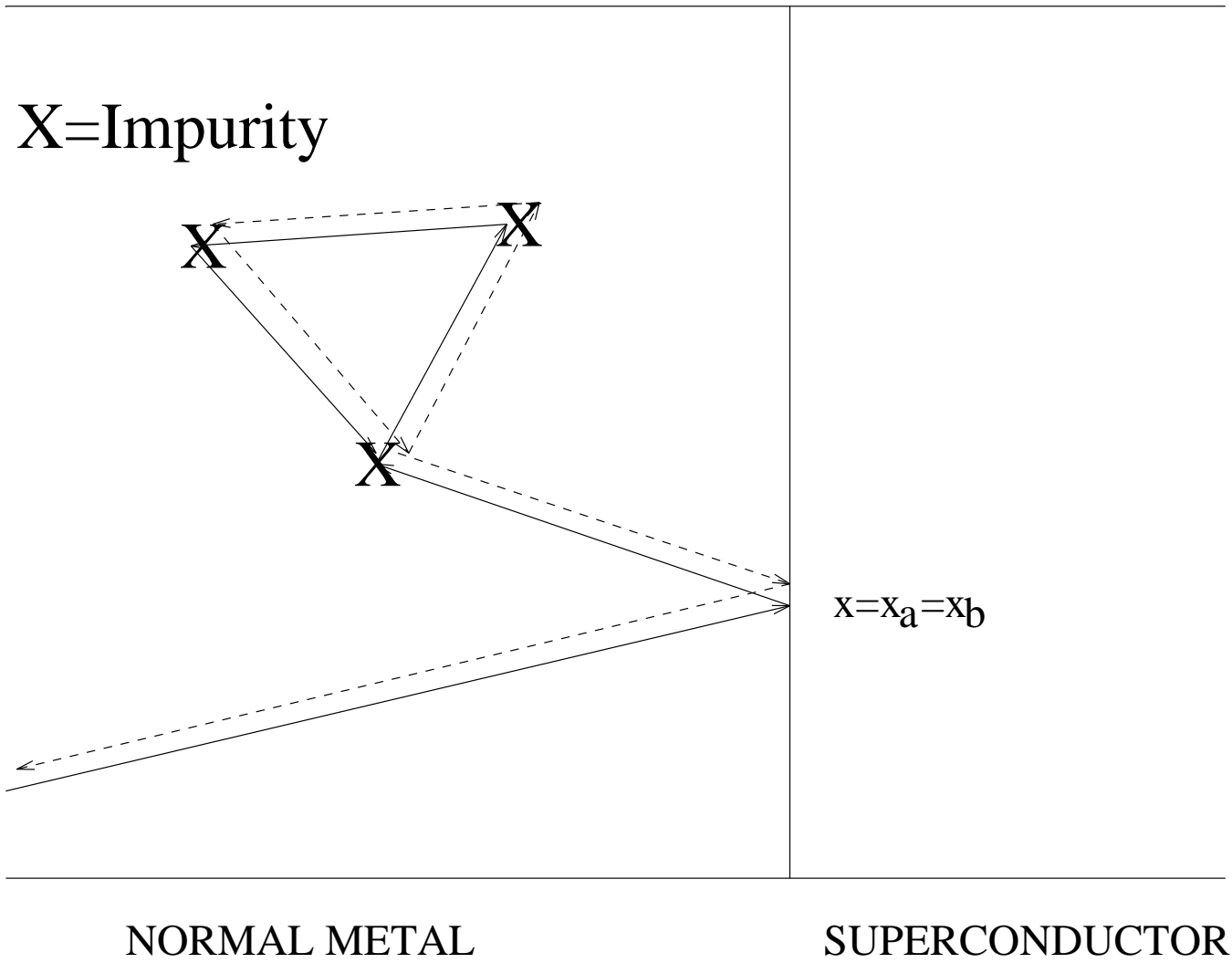
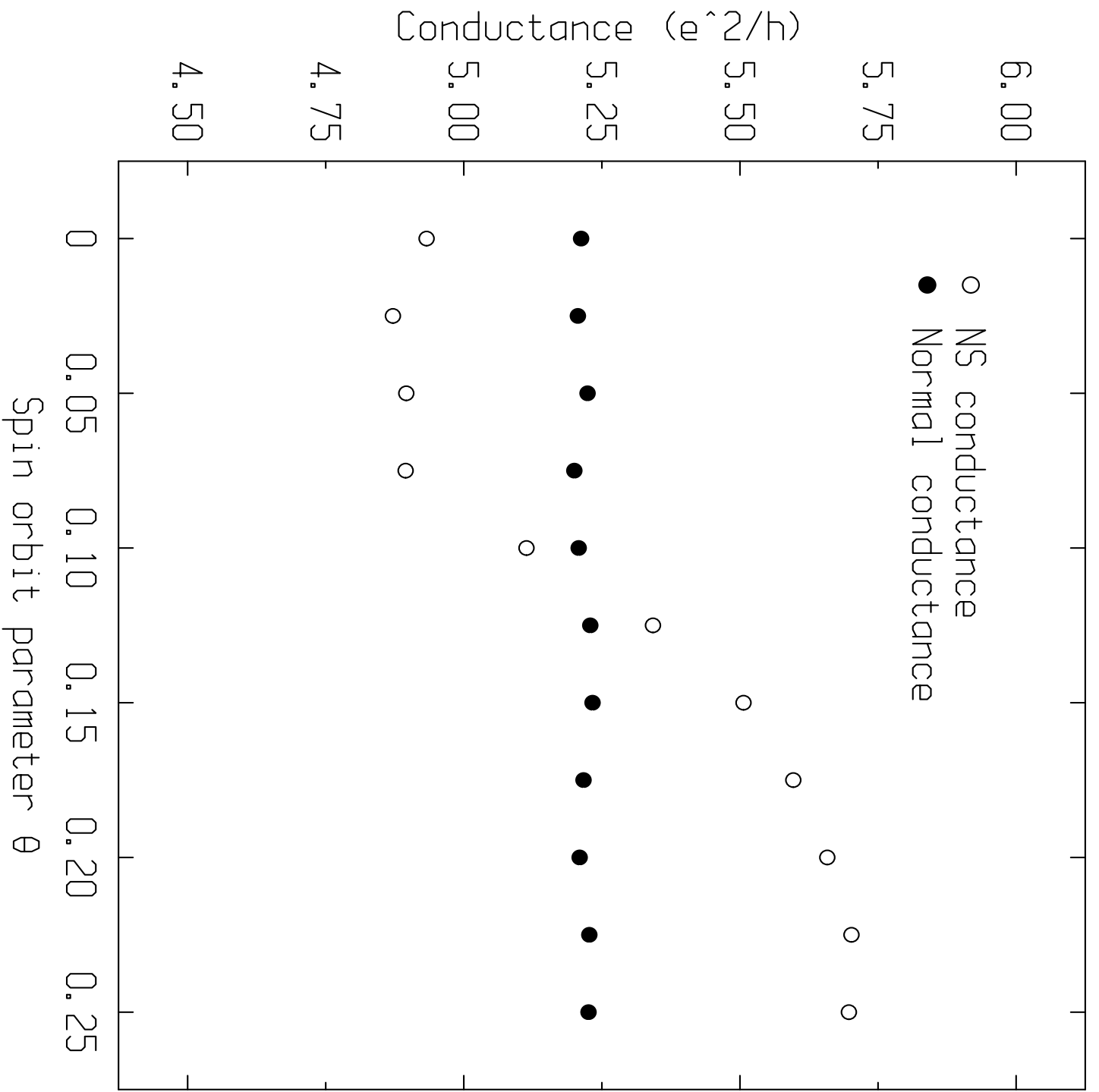
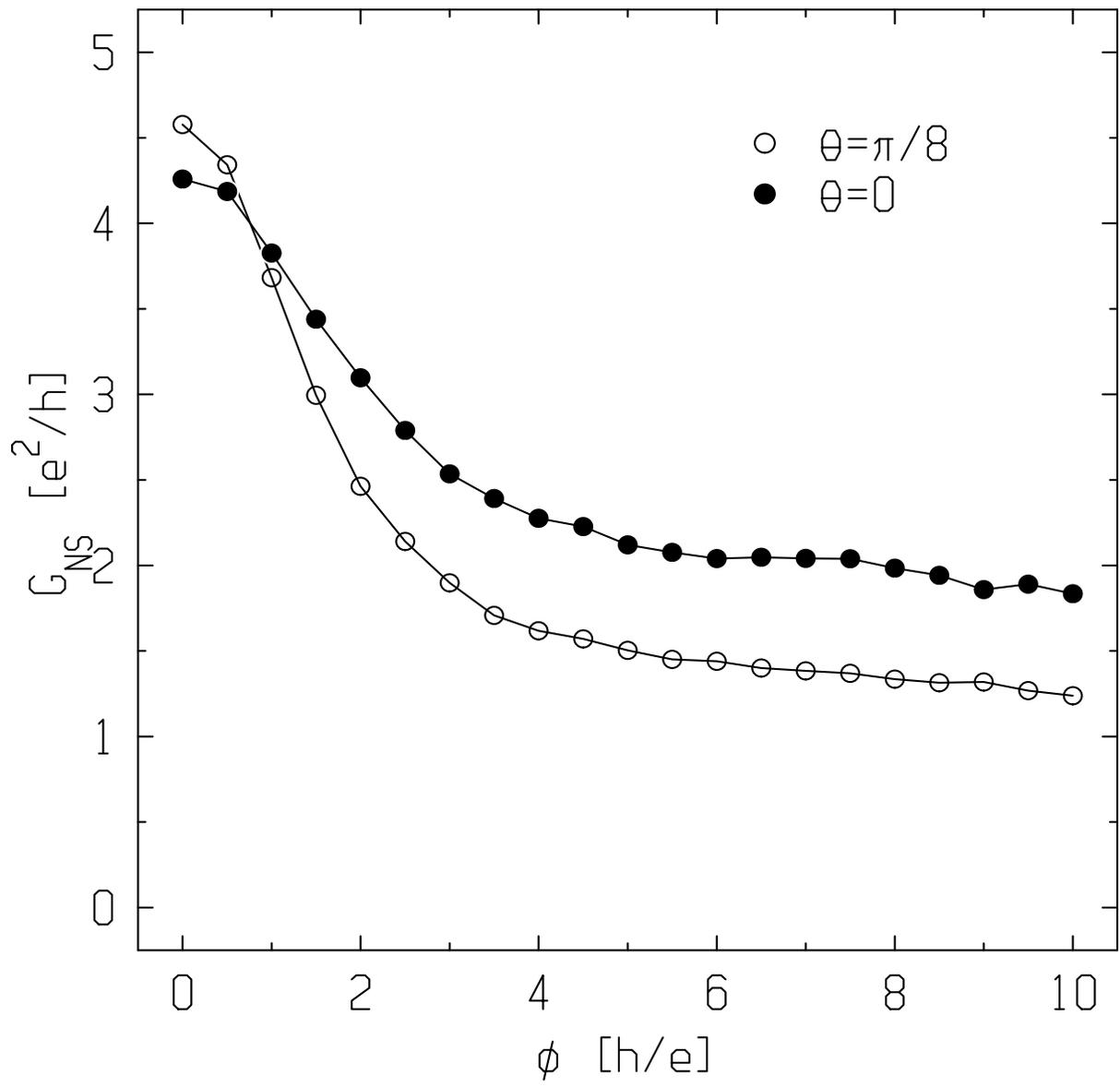
\end{document}